\documentclass[pre,twocolumn,showpacs]{revtex4-1}

\newcommand{\taupg}{\expt{\tau_p}}
\newcommand{\Cforce}{C_\mathrm{force}}

\newcommand{\Eq}[1]{Eq.~(\ref{eq:#1})}

\newcommand{\f}{\mathbf{f}}
\newcommand{\Fig}[1]{Fig.~\ref{fig:#1}}
\newcommand{\Figure}[1]{Figure~\ref{fig:#1}}

\newcommand{\REF}[1]{Ref.~\cite{#1}}
\newcommand{\NIB}{Nishikawa_Ikeda_Berthier:2021}

\newcommand{\Sec}[1]{Sec.~\ref{sec:#1}}
\newcommand{\App}[1]{Appendix~\ref{sec:#1}}
\renewcommand{\f}{\mathbf{f}}

\renewcommand{\v}{\mathbf{v}}

\newcommand{\expt}[1]{\left< #1\right>}
\newcommand{\gdot}{\dot{\gamma}}

\newcommand{\xiforce}{\xi_\mathrm{force}}

\usepackage{graphicx}
\usepackage{amssymb}
\usepackage{bm}

\begin{document}
\title{Relaxation times, rheology, and finite size effects}

\author{Peter Olsson}
\email{Peter.Olsson@tp.umu.se}

\affiliation{Department of Physics, Ume\aa\ University, 
  901 87 Ume\aa, Sweden}

\date{\today}   

\begin{abstract}
  We carry out overdamped simulations in a simple model of jamming---a collection of
  bi-disperse soft core frictionless disks in two dimensions---with the aim to explore the
  finite size dependence of different quantities, both the relaxation time obtained from
  the relaxation of the energy and the pressure-equivalent of the shear viscosity. The
  motivation for the paper is the observation [Nishikawa et al., J. Stat. Phys,
  \textbf{182}, 37 (2021)] that there are finite size effects in the relaxation time,
  $\tau$, that give problems in the determination of the critical divergence, and the
  claim that this is due to a finite size dependence, $\tau\sim\ln N$, which makes $\tau$
  an ill-defined quantity. Beside analyses to determine the relaxation time for the whole
  system we determine particle relaxation times which allow us to determine both
  histograms of particle relaxation times and the average particle relaxation times---two
  quantities that are very useful for the analyses. The starting configurations for the
  relaxation simulations are of two different kinds: completely random or taken from
  steady shearing simulations, and we find that the difference between these two cases
  are bigger than previously noted and that the observed problems in the determination of
  the critical divergence obtained when starting from random configurations are not
  present when instead starting the relaxations from shearing configurations. We also
  argue that the the effect that causes the $\ln N$-dependence is not as problematic as
  asserted. When it comes to the finite size dependence of the pressure-equivalent of the
  shear viscosity we find that our data don't give support for the claimed strong finite
  size dependence, but also that the finite-size dependence is at odds with what one would
  normally expect for a system with a diverging correlation length, and that this calls
  for a novel understanding of the phenomenon of shear-driven jamming.
\end{abstract}

\pacs{63.50.Lm,	
  45.70.-n	
  83.10.Rs 	
}
\maketitle

\section{Introduction}

The jamming transition is a zero-temperature phase transition from a fluid at low density
to a disordered solid at a larger density \cite{Liu_Nagel}. Much work on the jamming
transition has centered on static packings produced by starting from random configurations
and moving the particles to relax the energy related to the particle overlaps and thus
generate packings with vanishing overlap \cite{OHern_Silbert_Liu_Nagel:2003}. Another
path towards jamming is through shearing simulations \cite{Durian:1995}, which are usually
done at constant (low) shear rates and one then finds that the shear viscosity exhibits a
critical divergence as the jamming density is approached from below
\cite{Olsson_Teitel:jamming, Andreotti:2012}.

The determination of the critical exponent of this transition---the shear-driven jamming
transition---has however turned out to be a difficult task, both in experiments and in
simulations \cite{Olsson_Teitel:jamming, Hatano:2008, Heussinger_Barrat:2009, Hatano:2009,
  Hatano:2010, Tighe_WRvSvH, Hatano:2011, Otsuki_Hayakawa:2011, Andreotti:2012}, and that
is for two reasons: First because of the difficulty to get reliable values of the
viscosity, $\eta(\phi)$ at densities $\phi$ close to jamming and, second, because $\phi_J$
is not known and the value of the exponent depends sensitively on $\phi_J$. A way to
circumvent this second difficulty and determine a critical divergence---albeit a different
one---is by instead examining the dependence on the average number of contacts, $z$, which
is advantageous since its value at jamming is known to be $z_c=2d$. This kind of approach
was pioneered by Lerner et al \cite{Lerner-PNAS:2012} in a work where they determined the
pressure of sheared hard disks below $\phi_J$ with an innovative simulation method. They
were then able to determine a critical exponent from the slope of $\eta_p\equiv p/\gdot$
where $p$ is pressure and $\gdot$ is the shear strain rate ($\eta_p$ is just $p$ in their
notation) vs $\delta z\equiv z_c-z$. By furthermore examining the vibrational modes it was
clear that there is a unique lowest vibrational mode with frequency $\omega_\mathrm{min}$,
which is directly related to $\eta_p$. It turns out that this mode spans the whole system
and is closely related to the shearing.

A limitation of the method of \REF{Lerner-PNAS:2012} is that a matrix equation has to be
solved every time the contact network is changing, which is something that happens more
frequently as the simulations are performed closer to jamming. A different way to get
similar non-overlapping configurations is by doing two-step simulations
\cite{Olsson:jam-tau}: First running ordinary shearing simulations of soft particles and,
second, performing relaxation simulations from such starting configurations. For each such
simulation the relaxation time was determined from the exponential decay of energy (or
pressure, as in \REF{Olsson:jam-tau}) to the non-overlapping limit. This relaxation time
$\tau$, determined in overdamped dynamics, turns out to be directly related to
$1/\omega^2_\mathrm{min}$ in a Newtonian dynamics \cite{Lerner-PNAS:2012,
  Olsson:jam-tau}. These quantities are also found to behave the same as $\eta_p$
\cite{Lerner-PNAS:2012, Olsson:jam-tau, Harukuni-logcorr:2020}. To be clear, the
determination of the relaxation time $\tau$ also suffers from problems at the approach to
$\phi_J$, as the simulation times are proportional to the relaxation times, which grow
rapidly as jamming is approached.

Another way to determine the critical behavior is through a scaling analysis of $\eta_p$
that generalizes the behavior $\eta_p\sim(\phi_J-\phi)^{-\beta}$, valid for hard
disks---or, equivalently, for soft disks in the limit $\gdot\to0$---to finite shear rates
\cite{Olsson_Teitel:jamming, Olsson_Teitel:gdot-scale}. There does however seem to be
weaknesses with all methods and for these scaling analyses it is the presence of large
corrections to scaling, that badly complicate the analyses.

Both these methods were used in the recent, unexpected, finding of different critical
behaviors in two and three dimensions \cite{Olsson:jam-3D}. This result has however been
questioned since it is difficult to reconcile with other results that suggest that the
upper critical dimension should be equal to two \cite{Wyart:2005,
  Goodrich:2012}---implying that the critical exponents should be the same for all
$D\geq2$, but possibly complicated by log corrections for $D=2$.

With the above two methods in different ways determining the divergence by making use of
shearing simulations, the report of similar results when instead relaxing from random
configurations \cite{Ikeda__Berthier:2020-relax} was quite unexpected as it suggested that
the shearing is not a prerequisite for getting a configuration with a well-defined
relaxation time. The picture was however complicated even more when another paper from the
same group \cite{\NIB} argued for problems with the analyses due to the neglect of a
finite size effect. The main complaint was that there is a finite size dependence in the
relaxation time, $\tau\sim\ln N$, which makes $\tau$ an ill-defined quantity which cannot
be used for determining the critical behavior. This finite size effect---which we here
call the NIB effect for Nishikawa, Ikeda, and Berthier---is that systems with linear sizes
bigger than a certain characteristic length, $L>\xi_\mathrm{force}$ (see below or in
\REF{\NIB} for the definition of $\xiforce$) split into islands that relax
independently. The final relaxation of the total energy is then governed by the island
with the largest relaxation time and the fact that a bigger system tends to have a bigger
``largest relaxation time'' leads to the system size dependence $\tau\sim\ln N$
\cite{\NIB}.

Of relevance for the determination of the critical behavior from a scaling analysis of
$\eta_p$, is a further claim in \REF{\NIB} that $\eta_p$ is plagued by severe finite size
effects and that only values of $\eta_p$ for systems with $L>\xiforce$, are to be
trusted. Since this length is a rapidly growing function of $\phi$, the claim casts doubts
on all attempts made hitherto to determine the critical divergence from $\eta_p$, as well
as the feasibility of such attempts for the future.

The present paper is a critical reexamination of \REF{\NIB} to see to what extent the
conclusions presented there hold for a closer analysis. The conclusions in \REF{\NIB} of
relevance for the present work are (i) that the relaxation time is ill-defined and cannot
be used for studying the transition, and (ii) that the viscosity obtained from accessible
system sizes are not useful since they suffer from severe finite size effects. We find
these conclusions to be unnecessarily pessimistic.

Our study is on simulations using two different protocols: ``shearing'' which means that
configurations created during steady shearing simulations at different constant shear
rates are used as starting points in relaxation simulations, performed with $\gdot=0$
\cite{Olsson:jam-tau}, and "random", which means that the initial configurations are
created by positioning the particles by random \cite{Ikeda__Berthier:2020-relax,
  \NIB}. Our findings may be summarized as follows: (i) There are big differences between
the results from the random case and the results from the shearing case, shown by the
finite size dependence of $\tau$ being much more pronounced in the random case. This
finite size effect is however often unrelated to the NIB effect mentioned above. (ii) We
determine the relaxation time of the individual particles, $\tau_p$, and use them first to
make histograms $H(\tau_p)$ which allow for direct studies of the NIB effect and second to
determine the average relaxation time, $\taupg$. It turns out that this is a quantity with
a well-defined thermodynamic limit which thus opens up for a determination of the critical
behavior. (iii) In our search for the origin of this additional finite size dependence we
then turn to the density fluctuations that are present in randomly generated
configurations before the relaxations, and we find that the finite size dependence in
these fluctuations, to some extent, survive into the relaxed configurations. We argue that
this is at the root of the finite size dependence in $\tau$ as well as in other
quantities. (iv) A finite size dependence of $\tau$ was found in \REF{\NIB} to invalidate
the attempts to determine the critical divergence. We confirm the same kind of effect also
at higher densities and larger system sizes for the random case but find that it is not
present in the shearing case, and argue that it is the large density fluctuations that lead
to relaxed systems that have problematic properties as evidenced in several different
quantities.  (v) We also examine the relation between island size and relaxation time
valid in the NIB region where the system splits up into different islands and find a
direct proportionality between these quantities. The same kind of relation is also
obtained analytically for a one-dimensional model.  Another simple analytical argument
gives a relation between island size and density and suggests $\tau\sim(\delta\phi)^{-2}$
in good agreement with the behavior of $\taupg$ well below jamming. (vi) We then turn to
determinations of the correlation $\Cforce$ and the associated length $\xiforce$
introduced in \REF{\NIB} and find that even though the height of the maxima of $\Cforce$
differ by more than an order of magnitude for our two different protocols, the length
$\xiforce$ is about the same.  (vii) We finally consider finite size effects on $\eta_p$
and find that the onset of the finite size effect takes place at the same system size for
different densities in the range $\phi=0.830$ through 0.838, which is at odds with the
general expectation for finite size effects in critical phenomena, and is in contrast to
the claim in \REF{\NIB} of very severe finite size effects in $\eta_p$.

The organization of the paper is as follows: In \Sec{model} we introduce the simulation
methods and the measured quantities, in \Sec{Results} we give results for the relaxation
simulations, in \Sec{etap} we present the finite size study of the pressure equivalent of
the viscosity, and in \Sec{Discuss} we summarize the results.

\section{Models and measured quantities}
\label{sec:model}

For the simulations we follow O'Hern \emph{et al.}\cite{OHern_Silbert_Liu_Nagel:2003} and
use a simple model of bi-disperse frictionless disks in two dimensions with equal numbers
of particles with two different radii in the ratio 1.4. Length is measured in units of the
diameter of the small particles, $d_s$. We use Lees-Edwards boundary
conditions \cite{Evans_Morriss} to introduce a time-dependent shear strain
$\gamma = t\gdot$. With $r_{ij}$ the distance between the centers of two particles and
$d_{ij}$ the sum of their radii, the relative overlap is $\delta_{ij} = 1 - r_{ij}/d_{ij}$
and the interaction between overlapping particles is
$V(r_{ij}) = \epsilon \delta_{ij}^2/2$; we take $\epsilon=1$. The force on particle $i$
from particle $j$ is $\f^\mathrm{el}_{ij} = -\nabla_i V(r_{ij})$, which gives the
magnitude $f^\mathrm{el}_{ij}=\epsilon\delta_{ij}$. The simulations are performed at zero
temperature.

We consider the interaction force $\f^\mathrm{el}_i = \sum_j \f^\mathrm{el}_{ij}$ where
the sum extends over all particles $j$ in contact with $i$. The simulations discussed here
have been done with the RD$_0$ (reservoir dissipation) model
\cite{Vagberg_Olsson_Teitel:BagnNewt} with the dissipating force
$\f^\mathrm{dis}_i = -k_d \v_i$ where $\v_i\equiv \v_i^\mathrm{tot}-y_i\gdot\hat x$ is the
non-affine velocity, i.e.\ the velocity with respect to a uniformly shearing velocity
field, $y_i\gdot\hat x$.  In the overdamped limit the equation of motion is
$\f^\mathrm{el}_i +\f^\mathrm{dis}_i = 0$ which becomes $\v_i = \f^\mathrm{el}_i/k_d$.  We
take $k_d=1$ and the time unit $\tau_0 = d_s^2 k_d/\epsilon=1$. The equations of motion
were integrated with the Heuns method with time step $\Delta t/\tau_0=0.2$.  We simulate
with $N=256$ through $1048576\approx10^6$ particles to study finite size effects.

To determine the relaxation time we run simulations as described above at zero temperature
and fixed $\gamma$ which leads to an energy decreasing down towards zero; the simulations
are aborted when the energy per particle is $E<10^{-20}$. The relaxation time is then
determined from the exponential decay of the energy per particle by fitting $E(t)$ to
\begin{equation}
  E(t) \sim e^{-t/\tau},\quad E(t)<10^{-17}.
  \label{eq:E-t}
\end{equation}
For each parameter set the data are based on 4 to typically 100 different relaxations, but
sometimes up to 1000. The lower number is for some of the simulations for systems with
$N\approx10^6$, where fluctuations are small. The bigger numbers are for the smaller $N$.

A key observation in \REF{\NIB} is that different regions of the system may exhibit
different decay rates. To study this phenomenon we introduce and examine the relaxation
time of the energy for different particles. The elastic energy related to particle $i$ is
$E_i=\frac 12\sum_j \epsilon\delta^2_{ij}/2$, where the sum is over all particles $j$ in
contact with $i$. The relaxation time for particle $i$, $\tau_p^{(i)}$, is then determined
from the final stage of the relaxation of $E_i$, from configurations stored during the
relaxation. These configurations are stored with a time interval $\Delta$, from 100 to
10~000 time units. The smaller $\Delta$ are necessary to determine the shorter relaxation
times that are present at lower densities.

Using $E_i(t) \sim e^{-t/\tau_p^{(i)}}$ and letting $t'$ be the largest time for which
$E_i(t)> 10^{-20}$, the particle relaxation time is determined from
\begin{equation}
  \tau_p^{(i)} = \frac{\Delta}{\ln[E_i(t'-\Delta)/E_i(t')]}.
  \label{eq:taup}
\end{equation}
From these relaxation times, calculated for all particles, we determine histograms
$H(\tau_p)$. There is then always a small fraction of particles with unreasonable
relaxation times because of unstable configurations e.g.\ because they have only a single
contact. These particles are skipped in the final calculation of averages.

For big systems and large relaxation times we have found that one can get artifacts in the
determinations of $\tau_p^{(i)}$ due to the finite precision in the double precision
numbers used for the positions. This is an effect both of the fact that fewer bits are
available for the fraction part when storing larger coordinate values, which leads to a
lower precision in the position variables, and the fact that the net force, $f_i$ is
typically a factor of $\tau$ (which may be close to a factor $10^5$) smaller than the
typical interparticle force, $f_{ij}$, which means that the interparticle forces need to
be obtained with high precision. To handle that problem the relaxations for the biggest
$N$ at the higher densities were run with a version of the simulation program which uses
two variables---for integer part and fraction---to store to position coordinates. The
artifacts mentioned above were eliminated in that program version.

To examine the different simulation protocols we study the density fluctuations in both
initial and relaxed configurations---for the different cases of random and shearing---and
how these fluctuations depend on the size of a circular region $R$. This is done by
measuring $A(R)$, which is the total amount of particle area that is inside a circle of
radius $R$, such that the particles on the boundary contribute with only the fraction of
their area which is inside the circle. These areas are measured for a large number of
different configurations and center points, and the variance is obtained from
\begin{equation}
  \sigma^2_A = \expt{A^2(R)} - \expt{A(R)}^2.
  \label{eq:sigmaA2}
\end{equation}
For initial random configurations and $R\ll L$ the quantity $\sigma_A/\sqrt{N_R}$, where
$N_R=(\pi R^2/V)N$ is the average number of particles inside $R$, approaches a constant,
as discussed in \App{binomial}, and we therefore find it convenient to use that quantity
rather than the density fluctuations. The relation to the density fluctuations, determined
in the context of hyperuniformity \cite{Wu_Olsson_Teitel}, is
$\mathrm{var}(\phi_R)=\sigma_A^2/(N_R\expt{a}\pi R^2)$, where
$\expt{a}=\pi(d_s^2+d_b^2)/8$ is the average particle area.

We also determine the correlation function $\Cforce$ \cite{\NIB} which measures to what
extent the net particle forces in the relaxed configurations are correlated \cite{Cforce},
\begin{equation}
  \label{eq:Cforce}
  \Cforce(r) = \frac{\expt{\frac 1 N \sum_{i<j} \f_i\cdot\f_j \delta(r-r_{ij})}}
  {\expt{\frac 1 N \sum_i \f_i\cdot\f_i}}.
\end{equation}

\section{Results}
\label{sec:Results}

\subsection{Comparison of random and shearing}
\label{sec:Comp_ran_shear}

\begin{figure}
  \includegraphics[width=8cm]{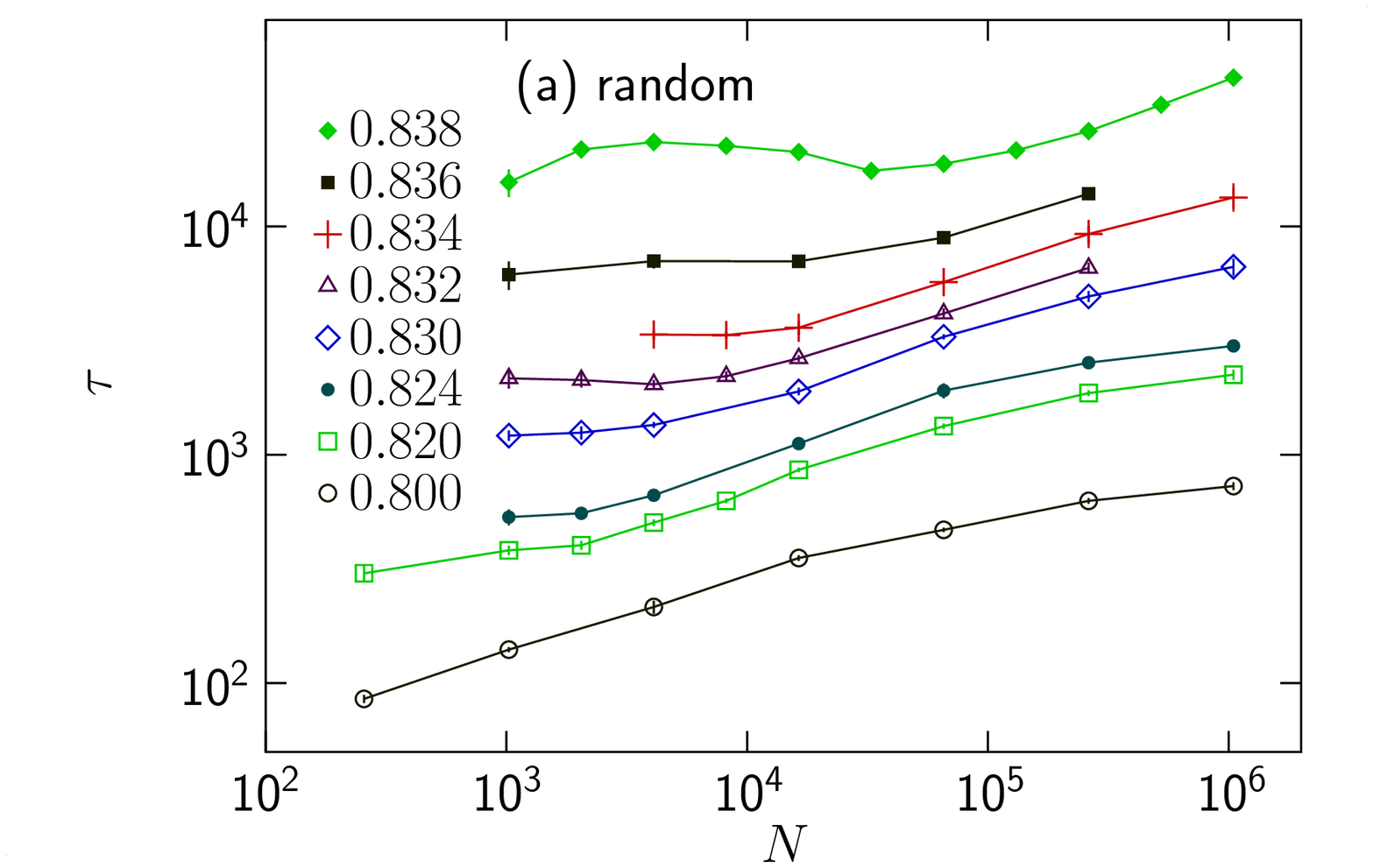}
  \includegraphics[width=8cm]{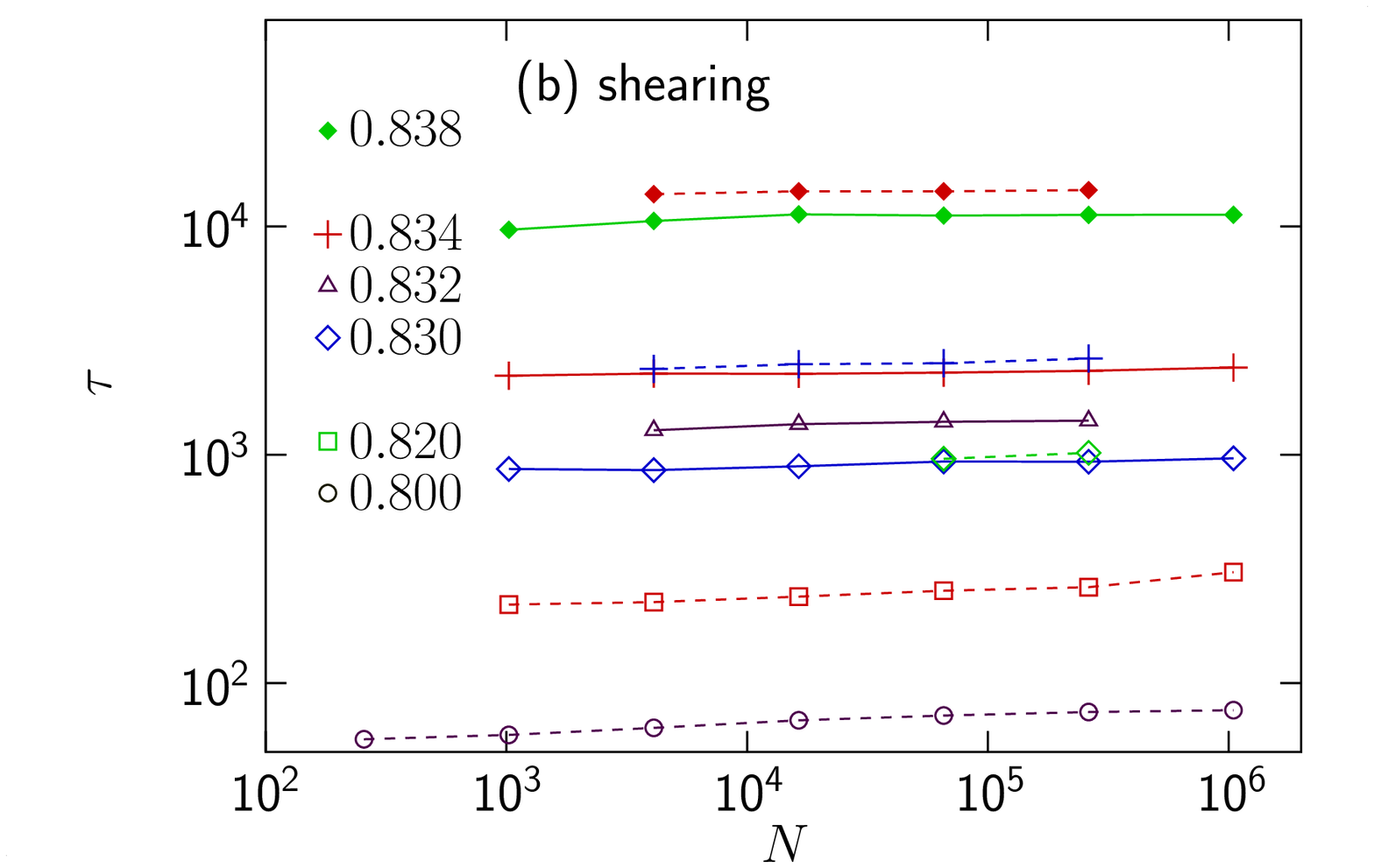}
  \caption{Finite size dependence of the relaxation time. The figures show the relaxation
    time from the exponential decay of the energy, \Eq{E-t}, versus number of particles
    for several different densities. Panel (a) is the behavior when starting from random
    configurations whereas panel (b) is the behavior when starting from configurations at
    steady shearing. The two data sets in panel (b) have been obtained when starting from
    configurations obtained with two different shear rates. Dashed lines connect data
    obtained starting from configurations generated with shear rate $\gdot=10^{-6}$, solid
    lines connect data for $\gdot=10^{-7}$. Note that the finite size dependence is
    considerably more pronounced when starting from random configurations.}
  \label{fig:tau-N}
\end{figure}

\Figure{tau-N} shows the relaxation times obtained at different $\phi=0.800$ through 0.838
and for a wide range of number of particles, $N$. Panel (a) shows results after relaxing
random configurations whereas panel (b) is after relaxing from shearing
configurations. For the shearing configurations there are data for initial shear strain
rates $\gdot=10^{-7}$ and $10^{-6}$, connected by solid and dashed lines,
respectively. When there is a shear strain rate dependence the relaxation time decreases
with decreasing initial shear strain rate \cite{Olsson:jam-tau}.

\Fig{tau-N}(a) does indeed underscore the message of \REF{\NIB} of significant finite size
effects in $\tau$. From a comparison of the two panels another clear message is the big
difference between $\tau$ from these different initial conditions.  The correlation times
obtained by relaxing random configurations are consistently larger and they also show much
more pronounced finite size effects.

The explanation in Ref.~\cite{\NIB} of the finite size effect is that the increase of
$\tau$ starts when the system is ``large enough to exhibit multiple correlated islands
where the relaxation dynamics can take place independently''. When that is the case it is
the island with the biggest relaxation time that will dominate the long-time behavior and
since a bigger system will have a larger number of different such islands it will be
expected to have a larger maximum relaxation time which, after a closer analysis, should
be $\tau\sim\ln N$ \cite{\NIB}.

\begin{figure}
  \includegraphics[width=8cm]{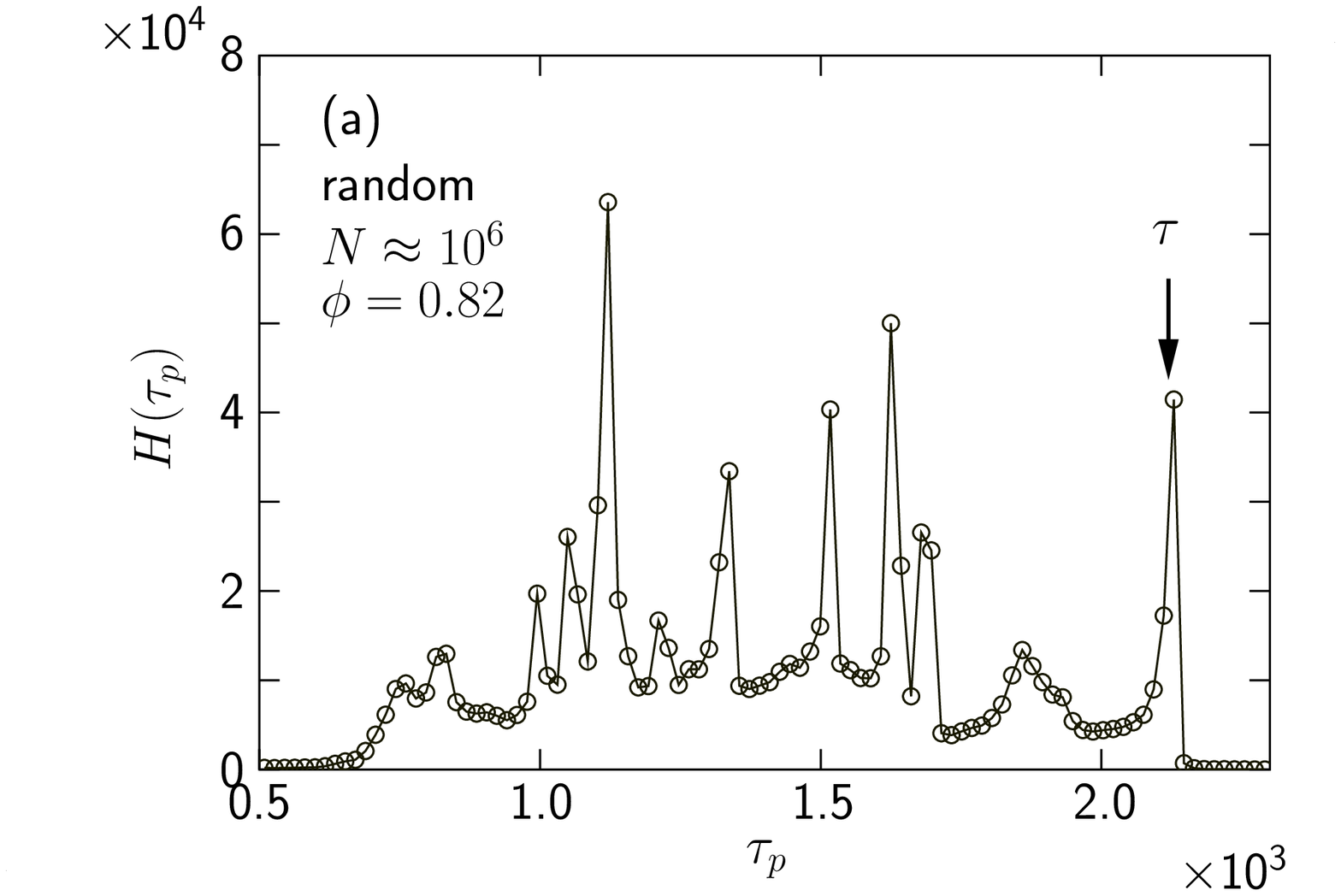}
  \includegraphics[bb=58 317 330 662, width=4.2cm]{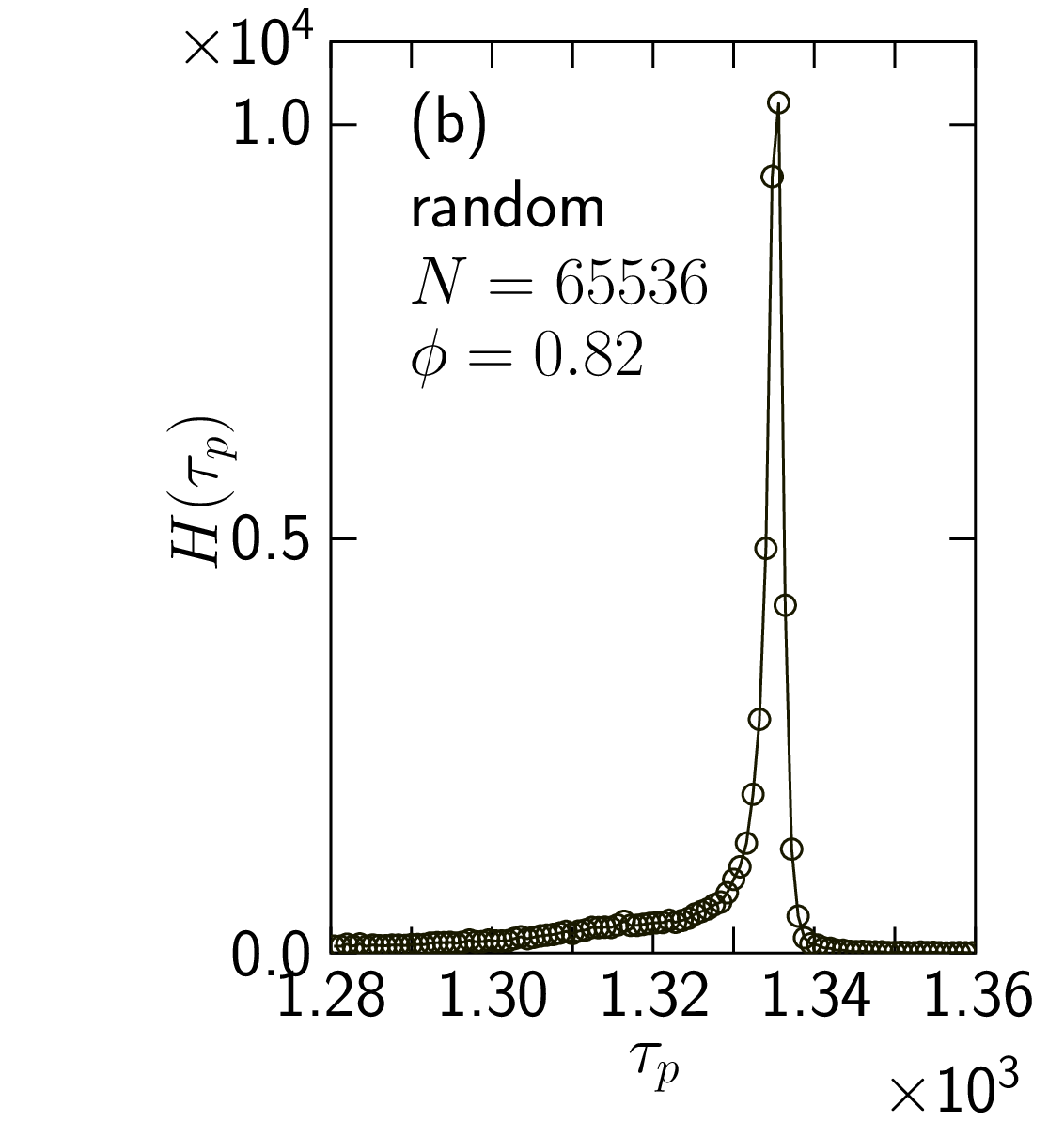}
  \includegraphics[bb=58 317 330 662, width=4.2cm]{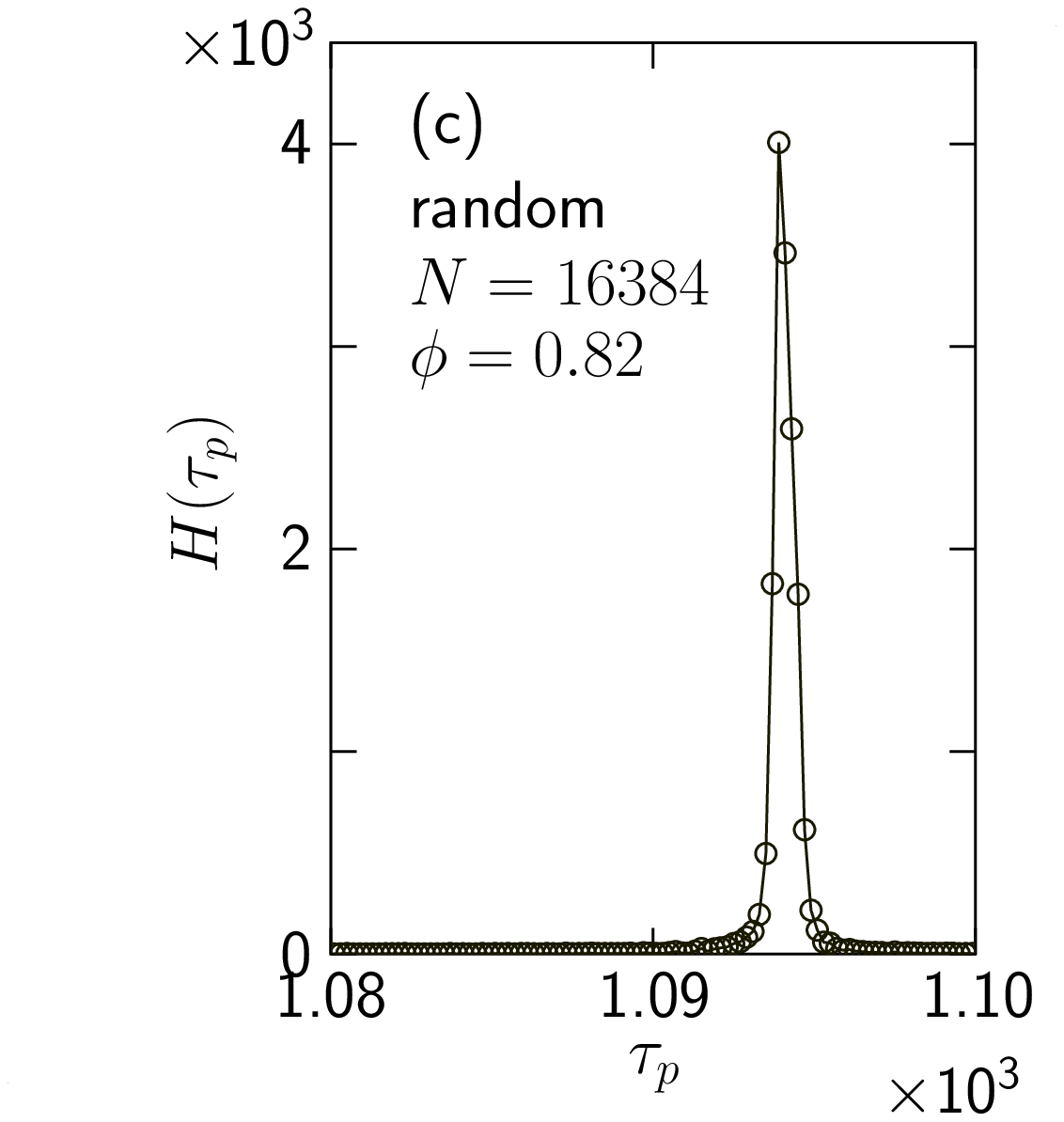}
  \caption{Histograms of particle relaxation times, $\tau_p$, produced by starting from
    random configurations at $\phi=0.820$ for two different system sizes. Panel (a) is for
    $N\approx10^6$ and gives clear evidence for the system splitting up in different parts
    with different relaxation times. (From the positions of the particles in the uppermost
    peak it is clear that that they indeed do belong to a localized island.) The arrow,
    which is $\tau$ from the decay of the total energy, illustrates that it is the slowest
    relaxation time that dominates this decay. As argued in \REF{\NIB} this is at the
    origin of the $\ln N$-dependence of $\tau$---here coined the NIB effect. In panel (c),
    which is for $N=16384$, all particles have the same relaxation time and there is no
    NIB effect, nevertheless, this is in a range of $N$ with a strong finite size
    dependence. The conclusion is that there are also other finite size effects beside the
    NIB effect. Panel (b) is for an intermediate number of particles, $N=65536$, and we
    here see a widening of the peak itself but also that the histogram extends to lower
    $\tau_p$ which is a precursor of some more peaks.  }
  \label{fig:H-taup-big-random}
\end{figure}

Though the explanation of this finite size effect---we call it the NIB effect from the
author names---appears to be correct we will argue that it cannot be the full story. For
more thorough investigations we determine the particle relaxation times $\tau_p$, as
described in \Eq{taup}, and construct histograms $H(\tau_p)$. The NIB effect is
illustrated with the histogram \Fig{H-taup-big-random}(a) for a single relaxation of a
system with $N=2^{20}\approx10^6$ particles at $\phi=0.820$. The peaks at different
$\tau_p$ are related to islands with different local relaxation times. The value of $\tau$
from the energy relaxation, $E\sim e^{-t/\tau}$, shown by the arrow, is close to the peak
at the highest $\tau_p$, which is consistent with the expectation that it is the slowest
particles that dominate the long time relaxation. We have verified that the particles in
this peak indeed do make up a localized island.

It is however clear that this is not the full explanation of the finite size effects in
\Fig{tau-N}. That figure shows that $\tau(N)$ is a monotonously increasing function of $N$
for $\phi=0.820$ and in the log-log plot the slope is actually the biggest around
$N\approx10^4$. Still, the histogram $H(\tau_p)$ from a single relaxation with $N=16384$
in \Fig{H-taup-big-random}(c) only gives evidence for a single peak and the same is true
for all our 16 examined realizations generated with the same parameters, $\phi=0.820$ and
$N=16384$. This makes it clear that there are also other effects at play in these systems,
beside the NIB effect. The histogram in \Fig{H-taup-big-random}(b) is for an intermediate
number of particles, $N=65536$, where the peak of the histogram is not quite as sharp and
one also sees a precursor, below the main peak, of the behavior in panel (a).

Though the histograms $H(\tau_p)$ open up for new possibilities for analyzing the systems,
some of the conclusions above may actually be arrived at from the information in
\REF{\NIB}. Since this relates to results throughout the paper that discussion is
relegated to \App{NIB-evidence}.

\begin{figure}
  \includegraphics[width=7cm]{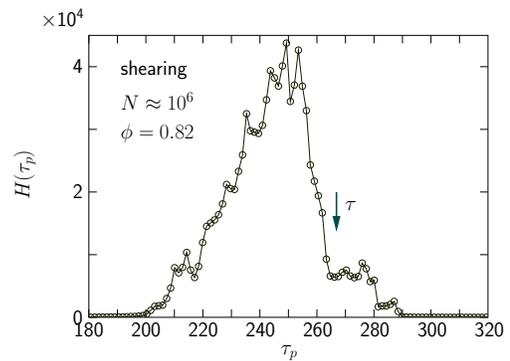}
  \caption{Histograms of particle relaxation times, $\tau_p$, produced by starting from
    shearing configurations with $N\approx10^6$, $\phi=0.820$, and $\gdot=10^{-6}$.}
  \label{fig:H-taup-big-shearing}
\end{figure}

To illustrate that there are both similarities and differences between the shearing and
the random cases, \Fig{H-taup-big-shearing} which is the same kind of figure as
\Fig{H-taup-big-random}(a), but for the shearing case, shows the histogram $H(\tau_p)$ for
the same parameters. There is again a clear signal of the NIB effect, but now as a broad
maximum rather than a number of well-separated peaks.

\subsection{Average particle relaxation time}
\label{sec:taupg}

With access to the particle relaxation times, $\tau_p$, for all individual particles it
becomes possible to determine the average $\tau_p$ as a different characterization of the
system. Since this is an estimate that gives equal weight to all particles and all regions
of the system one would expect this quantity not to be affected by the NIB effect. We use
the notation $\taupg$ though this quantity (in consistency with $\tau$ of
\cite{Olsson:jam-tau}) is determined as the geometric average,
$\expt{x}_\mathrm{geom} = \exp(\expt{\ln x})$, rather than an arithmetic average. The
different ways to determine the average do not give any noticeable differences for the larger
systems but, as discussed in \REF{Olsson:jam-tau}, the geometric average is a more
reasonable quantity for smaller systems.

\begin{figure}
  \includegraphics[width=8cm]{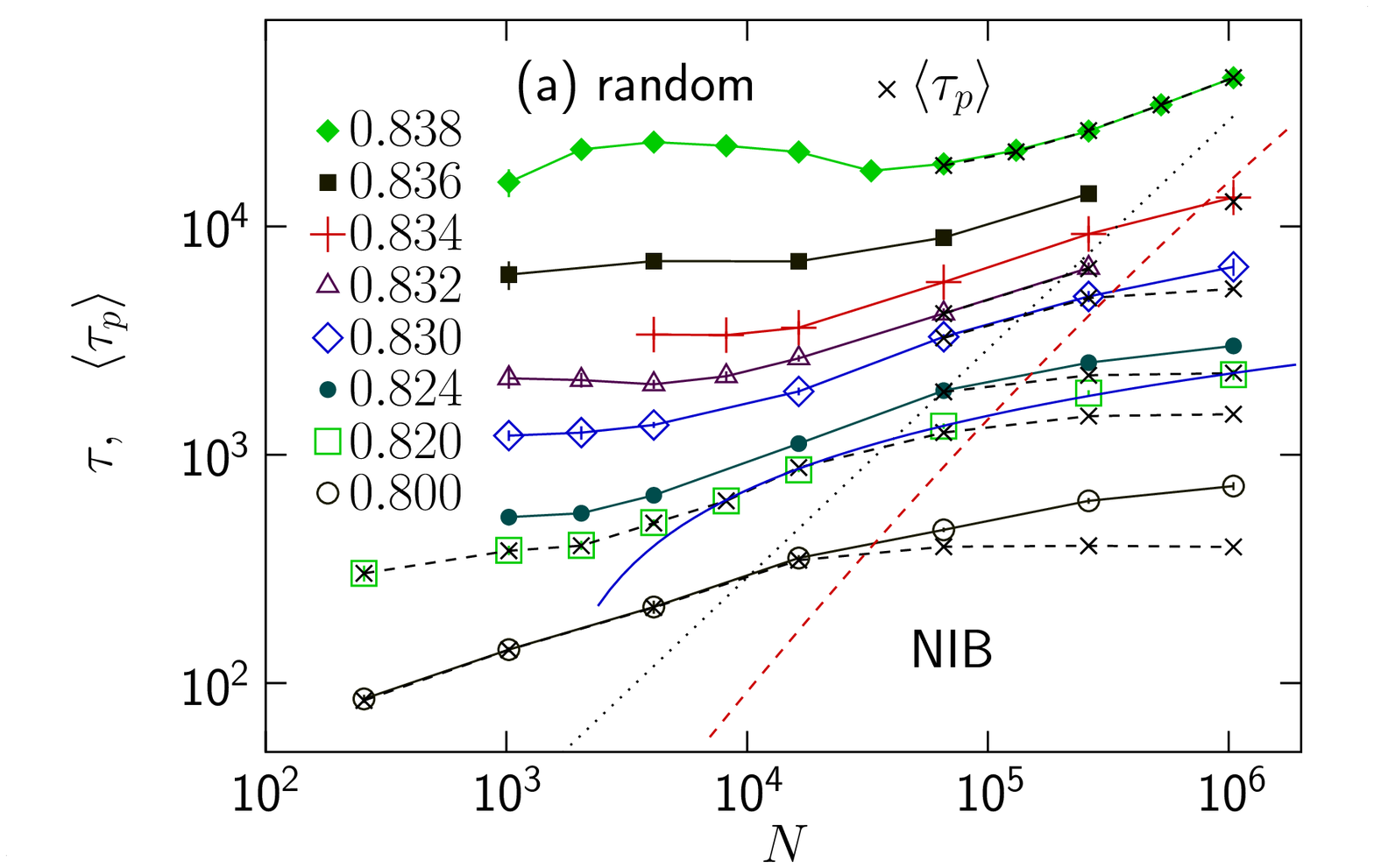}
  \includegraphics[width=8cm]{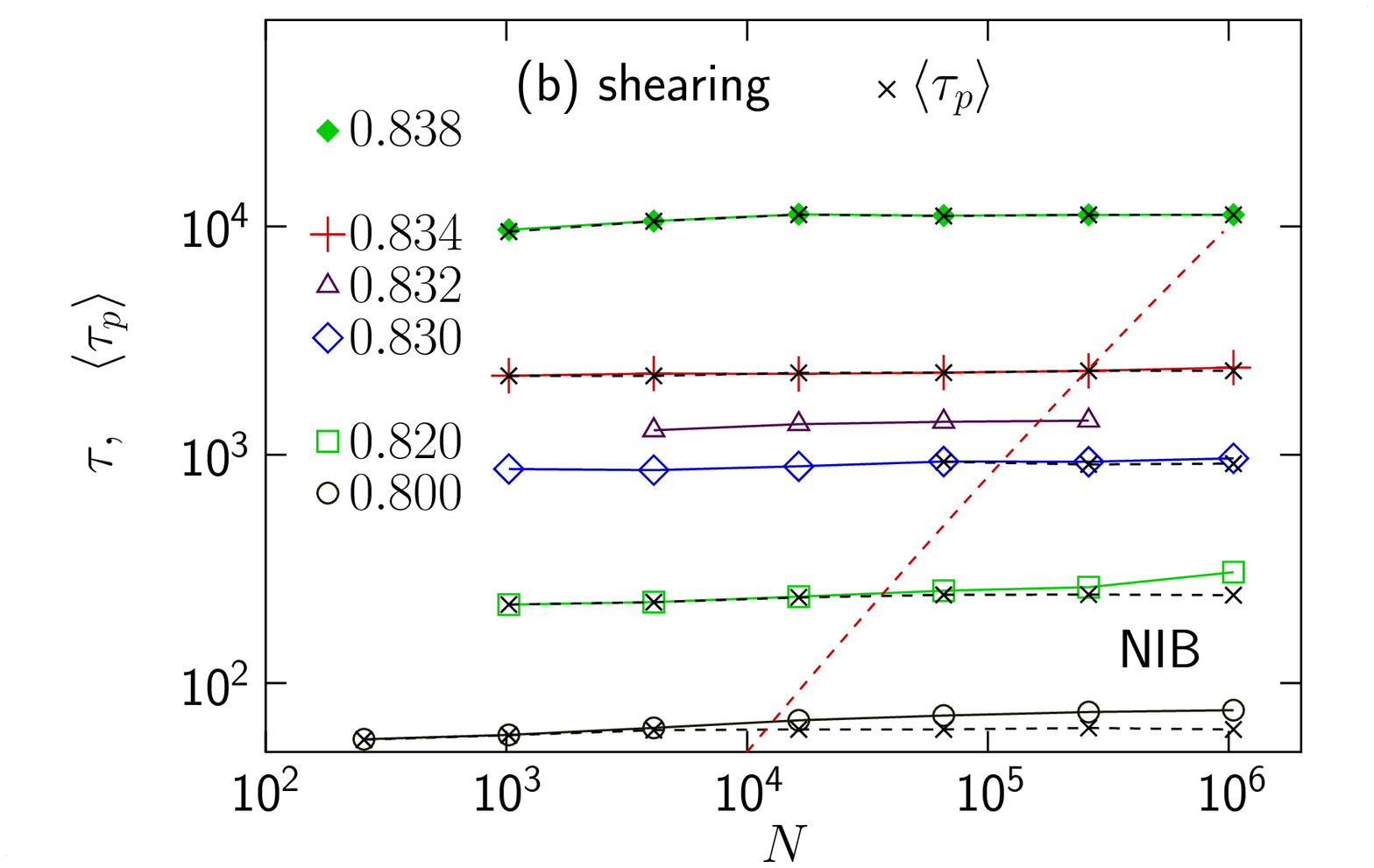}
  \caption{Finite size dependence of both the relaxation time, shown by different symbols
    for different $\phi$, and the average particle relaxation time, $\taupg$, shown by
    x:es. Just as in \Fig{tau-N} panel (a) is from random configurations whereas panel (b)
    is from shearing configurations. In each figure the dashed line is the approximate
    boundary of the NIB region, which is in the lower right part of the figures.  In turns
    out that $\taupg$ reaches finite limits for sufficiently large $N$, which is in
    contrast to $\tau$, which keeps increasing with increasing $N$. In both panels the
    region where $\tau$ and $\taupg$ differ is the NIB region; the boundary is indicated by
    the straight dashed line. The solid line in panel (a) through the data for
    $\phi=0.820$ (squares) shows that the data are well fitted to an $\ln N$-dependence
    even outside the NIB region. (The dotted line in panel (a) is the relation between
    $\tau_\mathrm{is}$ and $N_\mathrm{is}$ from \Fig{tauis_Nis}.)}
  \label{fig:tau-taup-N}
\end{figure}

\Figure{tau-taup-N} is $\taupg$ and $\tau$ vs $N$ for both random and shearing initial
configurations. The crosses show $\taupg$ whereas the other symbols are $\tau$ as in
\Fig{tau-N}. (To avoid cluttering the figure, panel (b) only shows the data for the lowest
$\gdot$ for each $\phi$.) In both figures $\taupg$ and $\tau$ mostly agree very well and
deviations are found only in the lower right corners of low $\phi$ and big $N$. The
difference between these two quantities is clearly due to the NIB effect and just as
mentioned above, $\taupg$ for each $\phi$ appears to approach a constant as $N\to\infty$
whereas $\tau$ continues to increase. A dashed line separates the NIB region from the
region where the finite size effect is of a different origin.

This finding leads to several conclusions. The first is that $\taupg$---in contrast to
$\tau$---is a quantity with a well-defined $N\to\infty$ limit, and that it may therefore
actually be used to determine the exponent of the critical divergence related to the
shear-driven jamming transition.

The second is that it now becomes possible to compare the relaxation time (in terms of
$\taupg$) for the two cases of random and shearing initial configurations. In the
large-$N$ limit it appears that $\taupg$ from random configurations is about a factor of
six bigger than $\taupg$ from shearing configurations.

A third conclusion regards the use of $\tau$ to determine the critical behavior from
shearing configurations. In \REF{\NIB} it was argued that $\tau$ may not be used to
determine the critical behavior for the jamming transition since it has no $N\to\infty$
limit. As argued above, $\taupg$ may instead be used for this purpose, but since the
determination of $\taupg$ is considerably more involved than the determination of $\tau$
it is interesting to note that $\tau$ actually may be used for the parameter values where
it may be considered a reliable estimate of $\taupg$, which happens to be the case for
most of the points in \Fig{tau-taup-N}(b) at the higher densities. This also applies to
the analyses in \REF{Olsson:jam-tau} which were done with data for higher densities,
$\phi\geq0.838$, only. (For comparisons with some of the figures in \REF{Olsson:jam-tau}
it should be noted that that paper shows $\tau$ versus $\delta z$ whereas the figures in
\REF{\NIB} and in \Sec{fss-shearing}, below, instead have $\delta z/2d$ on the $x$
axis. There is also a factor of two in difference in $\tau$ since it is there determined
from the the decay of pressure which depends on the overlap through $p\sim\delta$, whereas
$\tau$ is here determined from the energy, which is $\sim\delta^2$.)

A consequence of the NIB effect is the behavior $\tau\sim\ln N$ but that kind of behavior
of $\tau(N)$ for $\phi=0.820$ is---quite surprisingly---seen all the way down to $N=8192$,
which is quite far below the NIB region. This is the solid curve in
\Fig{tau-taup-N}(a). Though we cannot offer any real explanation, we will return to this
finding in conjunction with \Fig{saa-8200}(c) below.

The finding of larger relaxation times from random configurations than from shearing
configurations leads to questions about the origin of this difference in $\taupg$. It then
seems that the lower relaxation time is related to the larger $\delta z$ and thereby a
lower contact number, which in turn is a sign of a better relaxation. We thus believe that
it is the difference in the starting configurations---on the one hand random
configurations with huge density fluctuations and on the other configurations with a long
smoothing prehistory---that has big consequences for the final configurations. Further
evidence for this difference are given in the next section below.

\subsection{Density fluctuations}

We argued above that there is another finite size effect in the random case beside the NIB
effect, and we now turn to an attempt to understand its origin through an examination of
the size-dependence of the density fluctuations. We first examine the density
fluctuations in random initial configurations---the configurations before the relaxation
step---and show that they may be understood from elementary statistics. We then show that
these differences to some extent survive also to the relaxed configurations.

The basic mechanism follows from a consideration of $N_R$, which is the number of randomly
positioned particles that have their centers within a circle of radius $R$, which is a
fraction $f=\pi R^2/V$ of the total volume. This number is trivially on the average
$\expt{N_R}=fN$ and from the properties of the binomial distribution it follows that the
variance of the same quantity is $\sigma_{N_R}^2=f(1-f)N$. Taken together (and simplifying
the notation with $\expt{N_R}\rightarrow N_R$) this becomes
\begin{displaymath}
  \frac{\sigma^2_{N_R}}{N_R} = (1-f).
\end{displaymath}
We note that this expressions is a constant in the large-$N$ limit but decreases
$\propto R^2/N$ for finite $N$. This is thus the origin of the $N$ dependence in
$\sigma_A/\sqrt{N_R}$.

\begin{figure}
  \includegraphics[width=7cm]{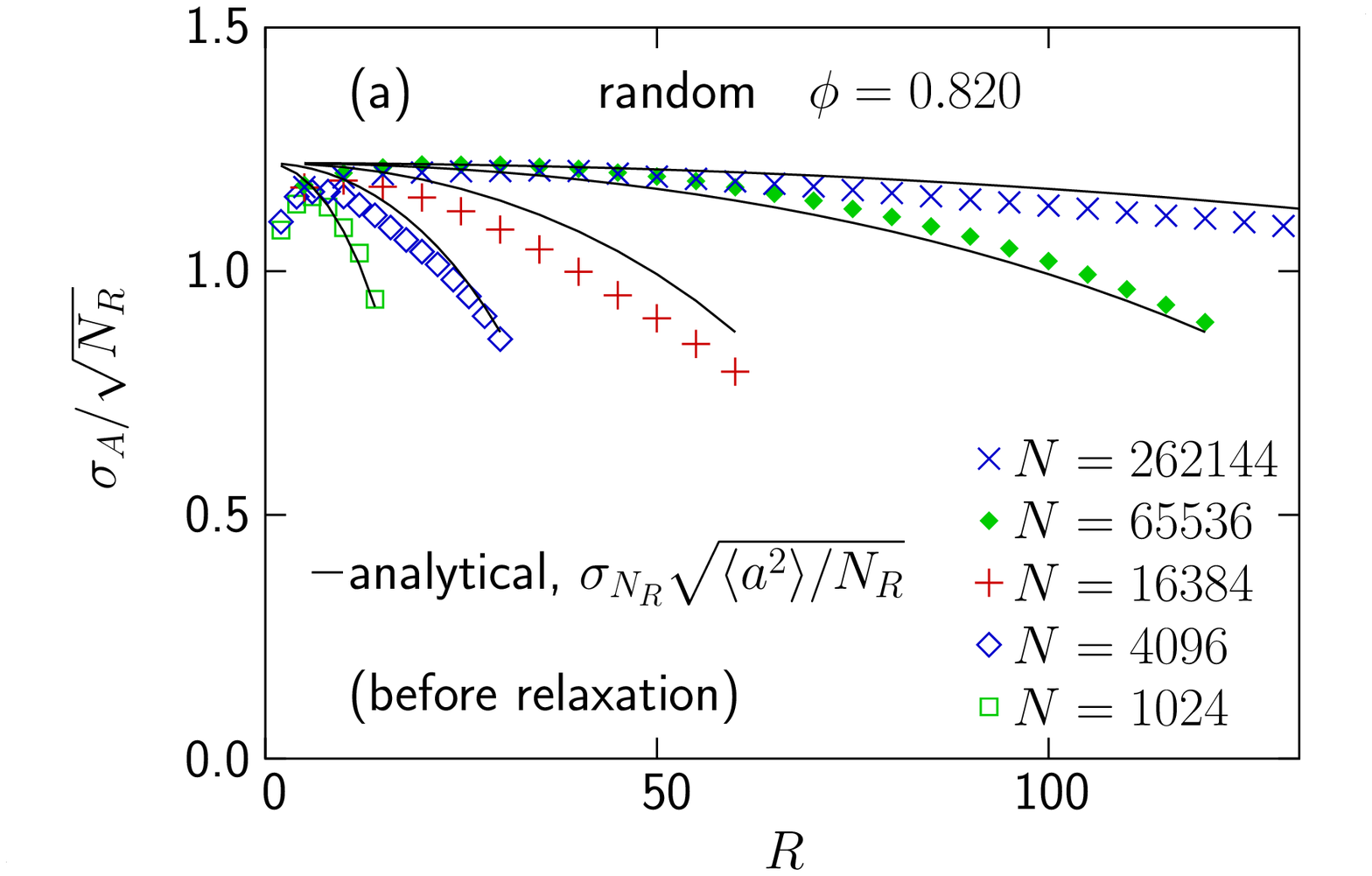}
  \includegraphics[width=7cm]{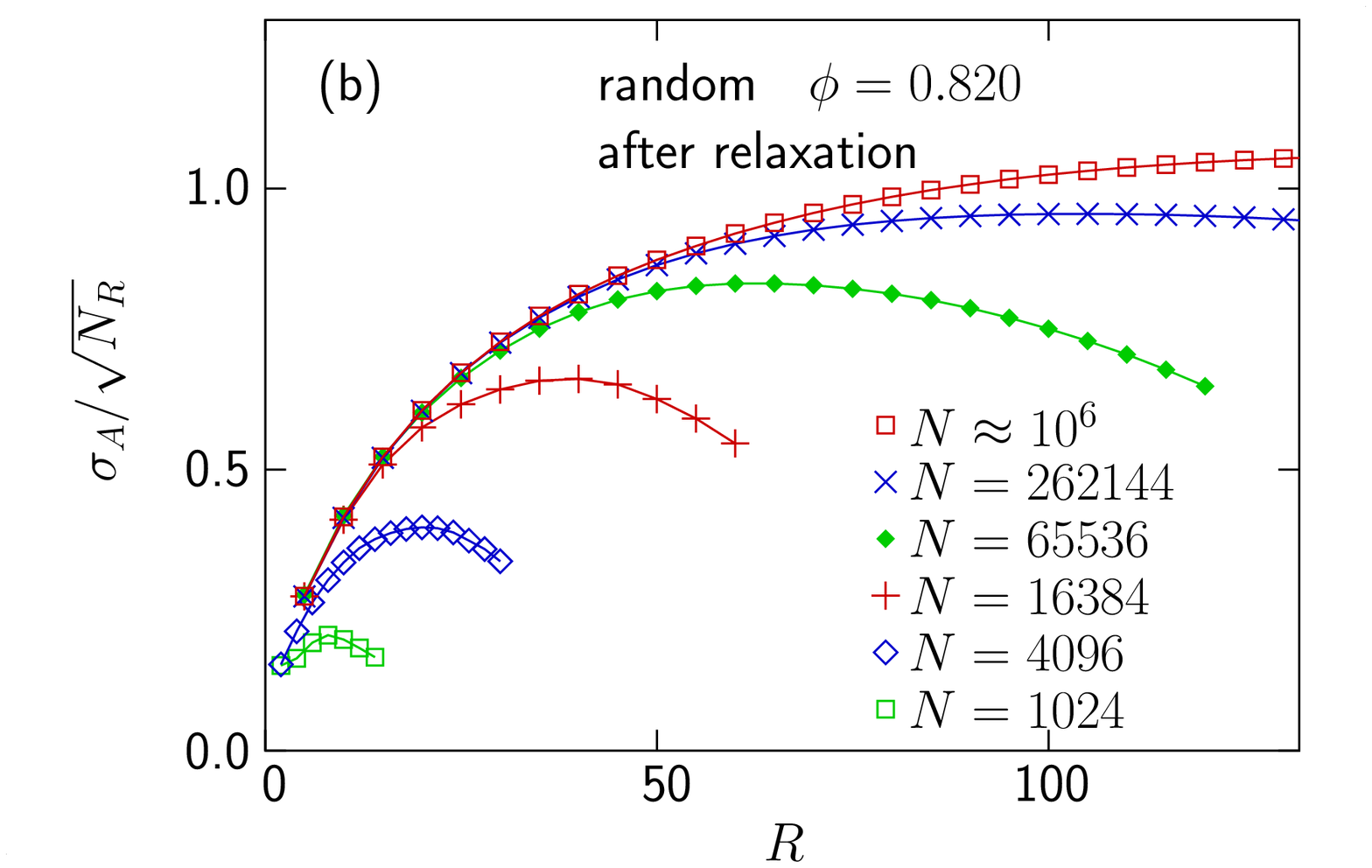}
  \includegraphics[bb=11 326 532 654, width=7cm]{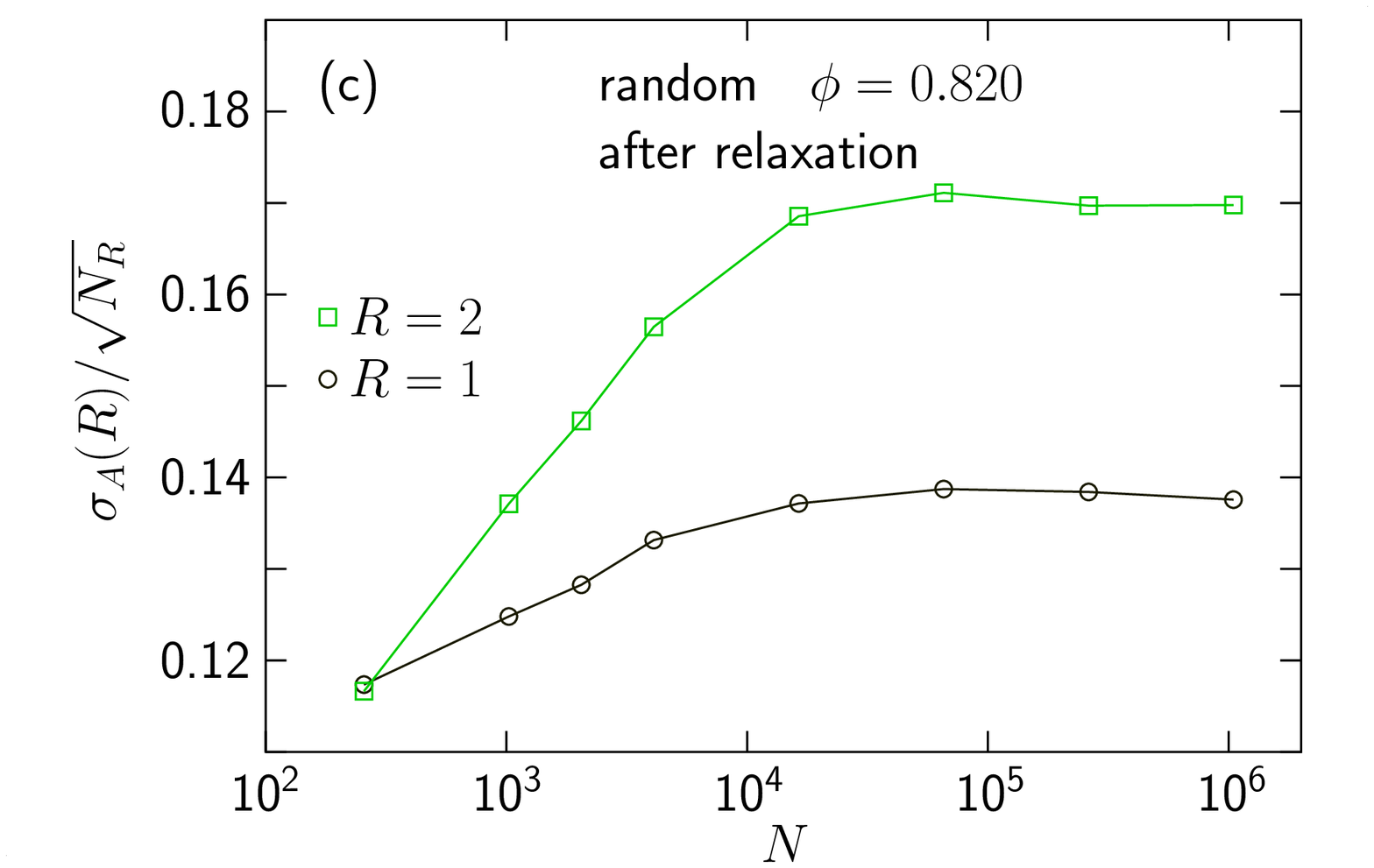}
  \caption{Density fluctuations for the random case at $\phi=0.820$ as measured by
    $\sigma_A/\sqrt{N_R}$. Panel (a) shows this quantity for initial configurations where
    the behavior is close to $\sigma_{A_c}/\sqrt{N_R}$, given by \Eq{sigma2Ac} and shown
    by solid lines. Panel (b), which is the same quantity after relaxation, shows that the
    finite size effects in the initial configurations to a large extent are present also
    in the relaxed configurations. (The lines are here just guides to the eye.)
    Panel (c) is the $N$-dependence of $\sigma_A(R)$ for $R=1$, 2. We see increases in
    $\sigma_A(r)$ up to $N\approx65536$ where the behavior approaches a constant value,
    and note that this is similar to the behavior of $\taupg$ at $\phi=0.820$ in
    \Fig{tau-taup-N}(a).
  }
  \label{fig:saa-8200}
\end{figure}

For analyzing our packings without overlaps we have found it convenient to measure $A(R)$,
and the related $\sigma^2_A$, from \Eq{sigmaA2}. The area $A(R)$ is the total particle
area inside a circle of radius $R$, such that the particles at the boundary contribute
with a fraction of their total area, according to definition I in
\REF{Wu_Olsson_Teitel}. A similar quantity which is related to $\sigma_{N_R}$ is obtained
by instead considering the area $A_c$ of the particles with their centers within this
circle, such that they always contribute with their total area---definition II in
\REF{Wu_Olsson_Teitel}.  Compared to $\sigma^2_{N_R}$ the variance of $A_c$ gets an extra
contribution from the different particle sizes, and as shown in \App{binomial} the
expression becomes
\begin{equation}
  \frac{\sigma^2_{A_c}}{N_R} = \expt{a^2}(1-f),
  \label{eq:sigma2Ac}
\end{equation}
where $\expt{a^2}$ is the average particle area squared,

\Figure{saa-8200}(a) shows the area fluctuations $\sigma_A/\sqrt{N_R}$ vs $R$, determined
with \Eq{sigmaA2} from random initial (unrelaxed) configurations for several different
$N$. The lines are for $\sigma_{A_c}/\sqrt{N_R}$ given by \Eq{sigma2Ac}. For the smallest
$R$, the data fall below the lines, as discussed in \App{binomial}, but otherwise the data
agree well with the prediction, considering that these are estimates of fluctuations,
which are more difficult to determine with high precision than are averages.

These random initial configurations with high energies and big particle overlaps are then
relaxed as discussed above and the area fluctuations from the relaxed configurations are
shown in \Fig{saa-8200}(b). We note that the density fluctuations for small $R$ are reduced the
most, but also that the finite size dependence from the initial configurations to some
extent survive into the relaxed configurations such that the area fluctuations are smaller
in the smaller systems.

\begin{figure}
  \includegraphics[width=7cm]{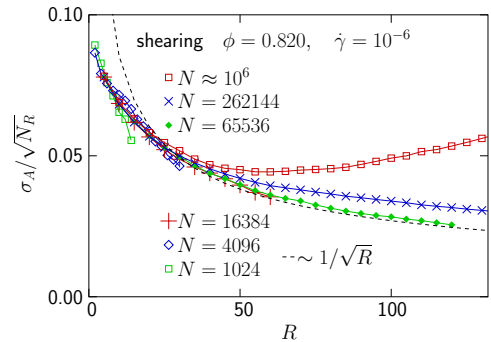}
  \caption{Density fluctuations for the shearing case at $\phi=0.820$ as measured by
    $\sigma_A/\sqrt{N_R}$. This shows that the density fluctuations are always small in
    configurations obtained with the shearing protocol. For system sizes $N\leq65536$,
    $\sigma_A/\sqrt{N_R}$ is essentially independent of $N$, but for $N\geq262144$---well
    into the NIB region---this quantity increases with increasing $N$, presumably because
    the system there splits up into different islands.  For comparison the dashed line is
    $\sim1/\sqrt R$, predicted from hyperuniversality, suggested at $\phi_J$.}
  \label{fig:saa-8200-shearing}
\end{figure}

The density fluctuations for the shearing case are shown in \Fig{saa-8200-shearing}. In
this case the fluctuations are considerably smaller---note the different scale. The reason
is clearly that the long shearing simulations before the relaxations give homogeneous
systems with small density fluctuations even in the unrelaxed systems. For $N\geq262144$,
in the NIB region, the density fluctuations are however somewhat bigger which is related
to the system splitting up into different islands with different local relaxation times
and presumably also different densities. We also note that hyperuniformity, suggested to
be present in jammed packings at $\phi_J$ \cite{Donev_ST:hyperuniform} would give
$\mathrm{var}(\phi_R)\sim(a+b\ln R)/R^3$ \cite{Wu_Olsson_Teitel} which translates to
$\sigma^2_A=(a+b\ln R)/R$. Though our data are far below jamming we find a similar
behavior, shown by the dashed line in \Fig{saa-8200-shearing}, which is
$\sigma_A\sim1/\sqrt R$, for the data below the NIB region.  (We also remark that the
density fluctuations, in the shearing case, are essentially the same before and after the
relaxation, since there are no big particle displacements during the relaxation.)

Since the density fluctuations are so wildly different for our two different simulation
protocols we believe that they are at the root of the different finite size dependencies of
$\tau$ in the two panels of \Fig{tau-taup-N}, and the effect that gives a large finite
size dependence for the random case will now be called the density fluctuation
effect---the DF effect.

A further link between the $N$-dependence of $\taupg$ in \Fig{tau-taup-N}(a) and the
density fluctuations is given by \Fig{saa-8200}(c) which shows the small $R$ behavior of
$\sigma_A(R)$ for $R=1$, 2. Similarly to the behavior of $\taupg$ we find that $\sigma_A$
increases at low $N$ and approaches constants at large $N$. The initial increase is
$\sigma_A \sim\ln R$, which is similar to the $\ln N$-behavior of $\tau(N)$ at
$\phi=0.820$, which was there found to extend far below the NIB region.

\subsection{Finite size effects in determinations of the critical exponent}
\label{sec:fss-shearing}

After introducing the particle relaxation times which gives methods for analyzing the NIB
effect, and identifying the DF effect, we are now ready to turn to examinations of the
critical divergence. From the relation $\tau\sim (\delta z)^{-\beta/u_z}$ it follows that
the exponent $\beta/u_z$ may be determined from the slope of $\tau$ vs $\delta z$ on a
double log plot \cite{Olsson:jam-tau}, but a test of this approach for the random case in
Fig.~1 of \REF{\NIB} revealed a problematic finite size dependence.

We will now argue that this is a problem for the random case only, and not for the shearing case. 

For the shearing case a close look at $\tau$ vs $N$ at $\phi=0.838$ in \Fig{tau-taup-N}(b)
shows that there is indeed a finite size dependence, and this observation could seem to
cast double on the use of $\tau$ for the determination of the critical
divergence. However, as already noted in \REF{Olsson:jam-tau} it turns out that the finite
size dependence of $\tau$ is accompanied by a similar finite size dependence of
$\delta z$, which make the points $(\tau,\delta z)$ for different $N$ fall on a common
curve.

\begin{figure}
  \includegraphics[width=7cm]{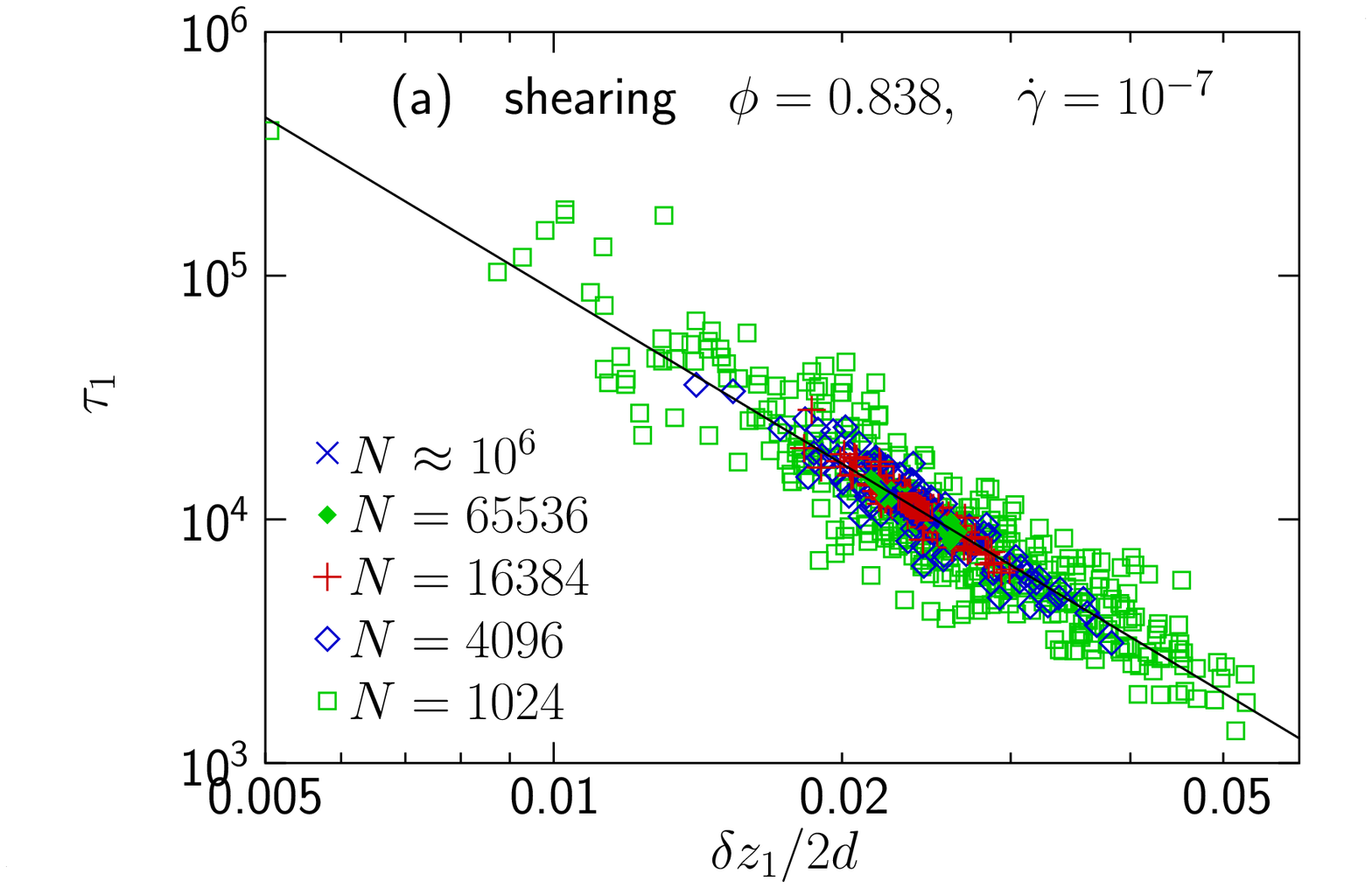}
  \includegraphics[bb=11 322 532 659, width=7cm]{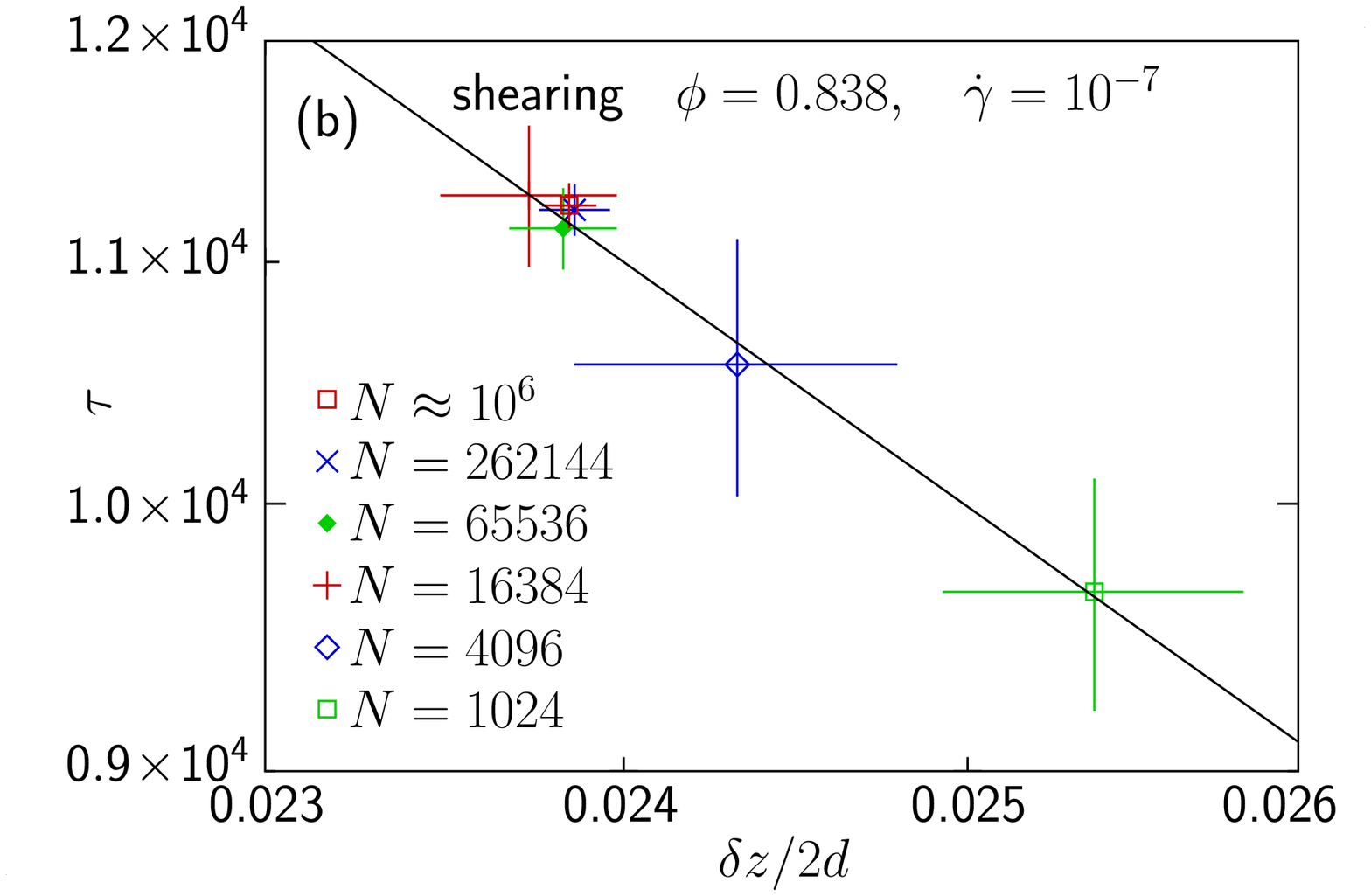}
  \caption{Relaxation time versus contact deficiency, $\delta$, for the shearing case
    and different system sizes, $N=1024$ through $\approx10^6$. Panel (a) shows individual
    measurements together with the line with slope $-b$, where $b=2.36$. Panel (b) shows
    $\tau$ vs $\delta z$, which are the geometric averages of the same quantities. This
    panel shows clearly that the finite size effects are small and unproblematic for the
    shearing case.}
  \label{fig:tau-dz-shearing}
\end{figure}

To illustrate this \Fig{tau-dz-shearing}(a) shows $\tau_1$ vs $\delta z_1/2d$ for
$\phi=0.838$ and several different sizes on a double-log scale. The subscript ``1''
signifies that these are values from single relaxations. The points for the bigger sizes
are close together whereas the points for smaller sizes spread considerably more
\cite{Olsson:jam-tau}. The data are suggestive of an algebraic behavior and a fit gives
the line governed by the exponent $-b$, with $b=2.36$. (For data closer to criticality,
i.e.\ smaller $\delta z$ and larger $\tau$, this exponent $b\to\beta/u_z\approx 2.7$
\cite{Olsson:jam-tau}; the value $b=2.36$ is only an effective exponent.) The finite size
dependence is too small to be clearly visible in \Fig{tau-dz-shearing}(a) and we therefore
calculate $\tau$ and $\delta z$ for the different $N$ as geometric averages of $\tau_1$
and $\delta z_1$ and plot $\tau$ vs $\delta z/2d$ in \Fig{tau-dz-shearing}(b). It then
turns out that the obtained points indeed do fall on the line in the figure, obtained from
the fit to the data in panel (a), and this shows clearly that the observed finite size
effect doesn't cause any problems for the determination of the divergence.

\begin{figure}
  \includegraphics[width=7cm]{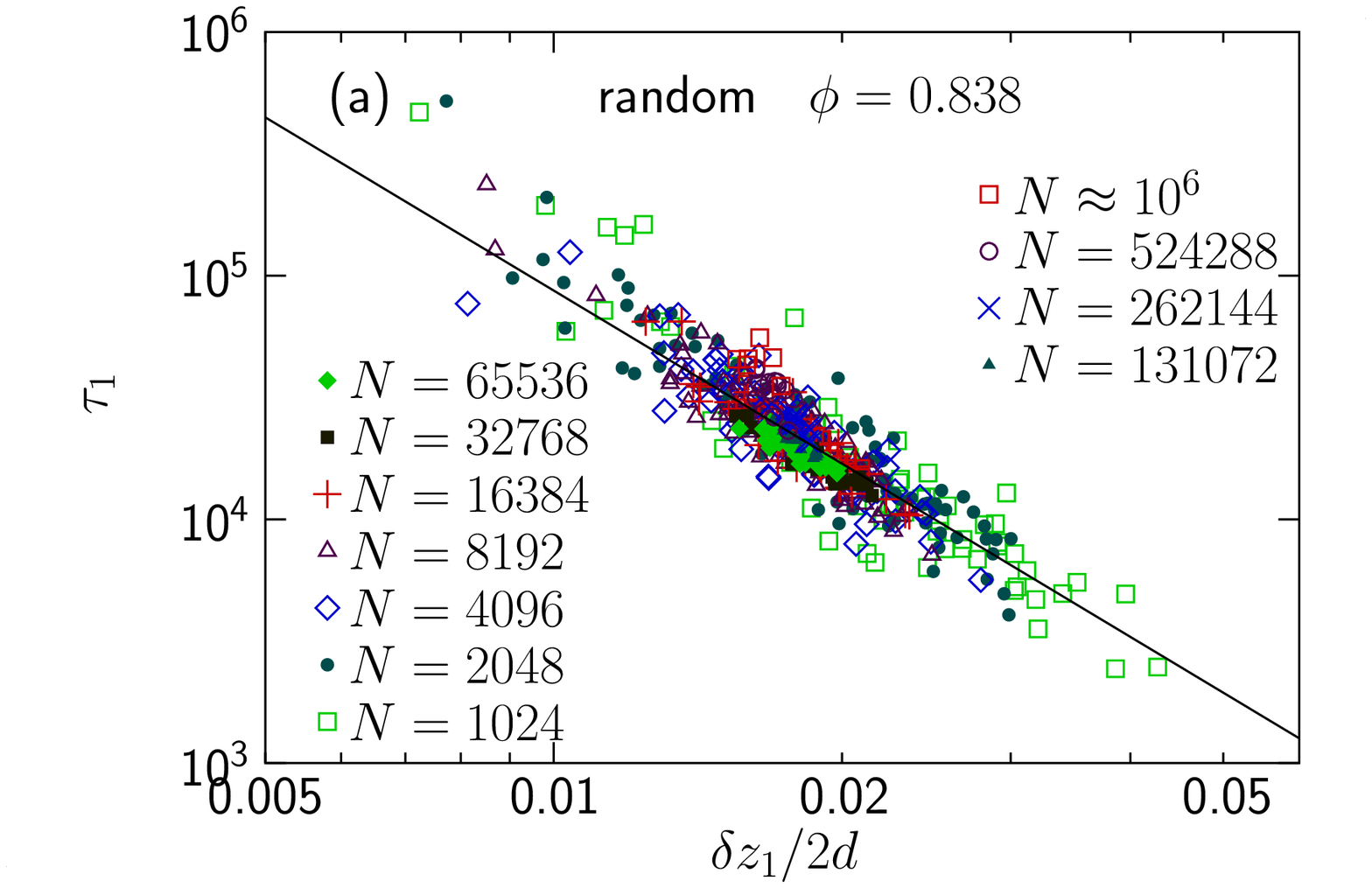}
  \includegraphics[bb=11 322 532 659, width=7cm]{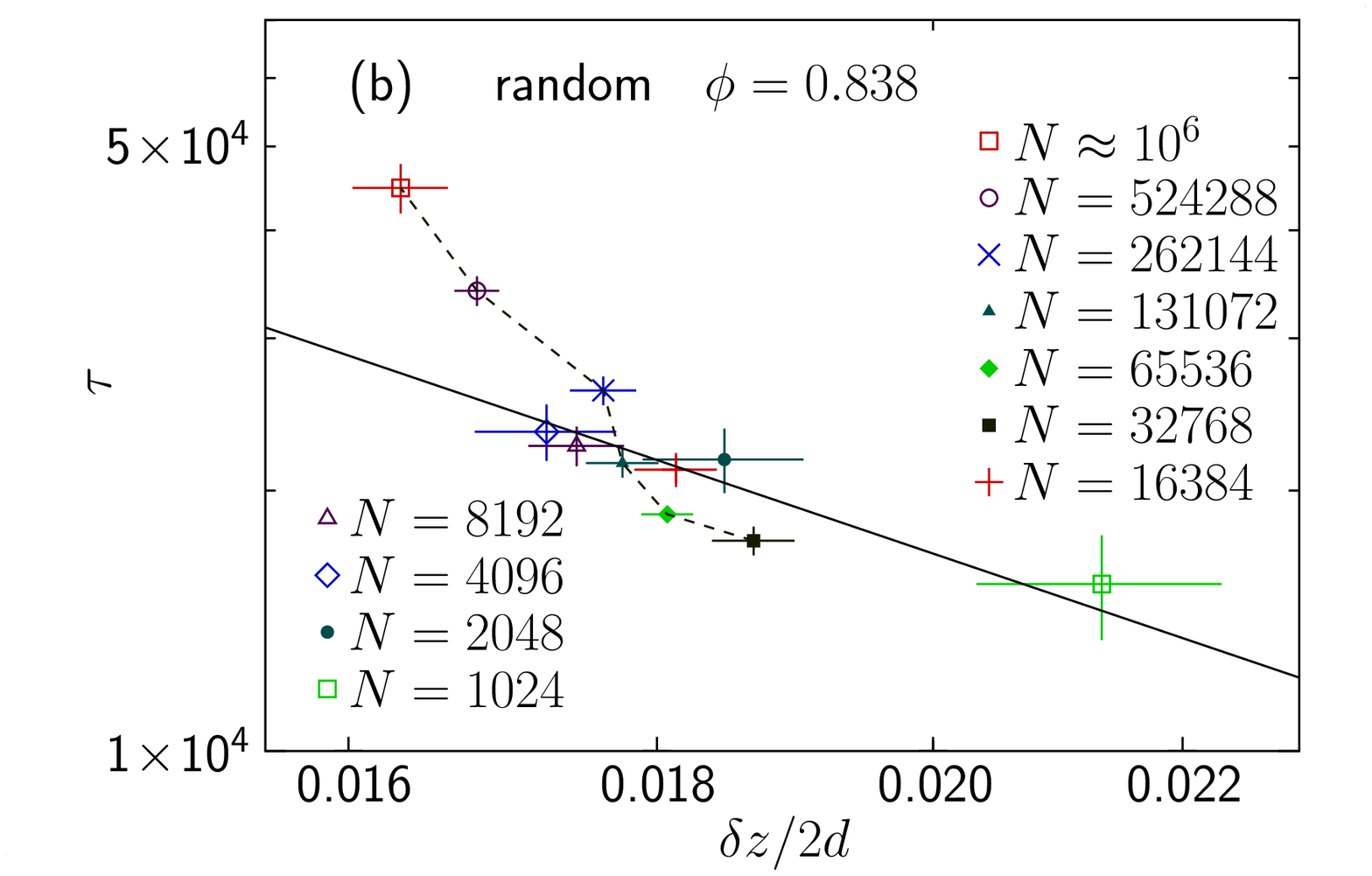}
  \caption{Relaxation time versus contact deficiency, as $\delta z/2d$, for the random
    case and different system sizes, $N=1024$ through $\approx10^6$. Panel (a) shows
    individual measurements. Also shown is the line from \Fig{tau-dz-shearing}(a), which
    is included to illustrate the ``universality'' suggested in
    \REF{Ikeda__Berthier:2020-relax}. To show the results in more detail panel (b) are
    the geometric averages of $\tau_1$ and $\delta z_1$. For sizes up to $N=16384$ seem to
    agree with the solid line (from \Fig{tau-dz-shearing}(a)) whereas the data for
    $N\geq32768$, connected by the dashed line, deviate clearly.}
  \label{fig:tau-dz-random}
\end{figure}

\begin{figure}
  \includegraphics[bb=11 322 532 659, width=7cm]{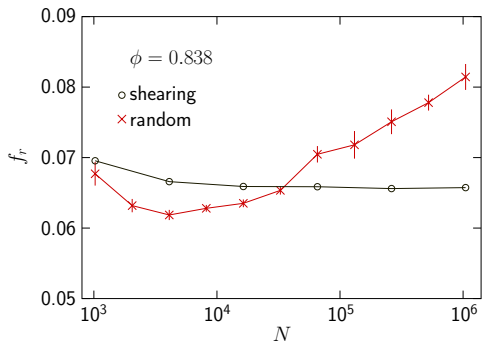}
  \caption{Fraction of rattlers at $\phi=0.838$, $N=1024$ through $N\approx10^6$ for both
    the random case and the shearing case. The failure of $f_r(N)$ to saturate for
    protocol random suggests a problem with this protocol.}
  \label{fig:frattler}
\end{figure}

For comparison, \Fig{tau-dz-random}(a) shows the same kind of data for random starting
configurations as in \Fig{tau-dz-shearing}(a), and the first observation is that two data
sets are very similar, especially since the solid line, which is taken from
\Fig{tau-dz-shearing}(a), fits well to the data. \Fig{tau-dz-random}(b) which give the
geometric averages, $\tau$ and $\delta z$, however, shows that there are also some
differences. Whereas data for $N\leq16384$ fit nicely to the line, the data for
$N\geq524288$ and $\approx 10^6$ are clearly off the line and it could be that there are
deviations from the expected behavior for all the data with $N\geq32768$---the points
connected by the dashed line. That this is not due to the NIB effect is clear since each
of the histograms of $\tau_p$ for our six different relaxations with $N\approx10^6$ and
$\phi=0.838$ have only a single peak.

The finite size dependence in \Fig{tau-dz-random} is the same kind of effect as was first
identified in Fig.~1 of \REF{\NIB} and was there attributed the the NIB effect. From
\Fig{tau-taup-N}(a) one may however, again, conclude that this cannot be explained by the
NIB effect, since the lowest $\tau$ for $N=262144$ in Fig.~1 of \REF{\NIB} is
$\tau\approx 5\times10^3$ whereas the NIB effect is only expected to be visible below
$\tau\approx3\times10^3$. We instead attribute this problematic finite size dependence to
the DF effect, which is consistent with this effect being visible for the random case and
not for the shearing case.

We have not been able to pinpoint the precise reason for the deviations from the expected
algebraic divergence for the random case, but we note that unexpected behaviors are
present in many different quantities. One such quantity that behaves strangely in the
random case is the fraction of rattlers, shown in \Fig{frattler} for both the shearing and
the random cases. For the shearing case $f_r$ is well-behaved and just decreases slowly
towards a constant as $N$ increases, but for the random case $f_r$ increases and fails to
saturate for our available system sizes. (Note that there is no direct relation between
$f_r$ and the contact number $z$; $f_r$ is related to the number of particles that are
eliminated before the calculation of $z$.) To us this suggests that the method to relax
randomly generated configurations is flawed as it leads to configuration with big density
fluctuations and spurious finite size dependencies. Surprising behaviors are indeed also
seen in \Fig{tau-taup-N}(a) where $\tau(N)$ is non-monotonic at $\phi=0.838$, and where
precursors of this non-monotonicity are also found at the lower densities.

The conclusion from this section---which is also one of the main conclusions of the
paper---is thus that it is the finite size dependence caused by the DF effect (density
fluctuations) that causes problems for the determination of the exponent $\beta/u_z$ in
the random case but that there are no such problems in the shearing case. We stress that
the data at high densities investigated here are not affected by the NIB effect, but we
also point out that one should watch out for this effect for large systems well below
$\phi_J$. In the next section we turn to the NIB region and properties of the individual
islands.

\subsection{Sizes of islands in the NIB region}

\begin{figure}
  \includegraphics[width=6cm]{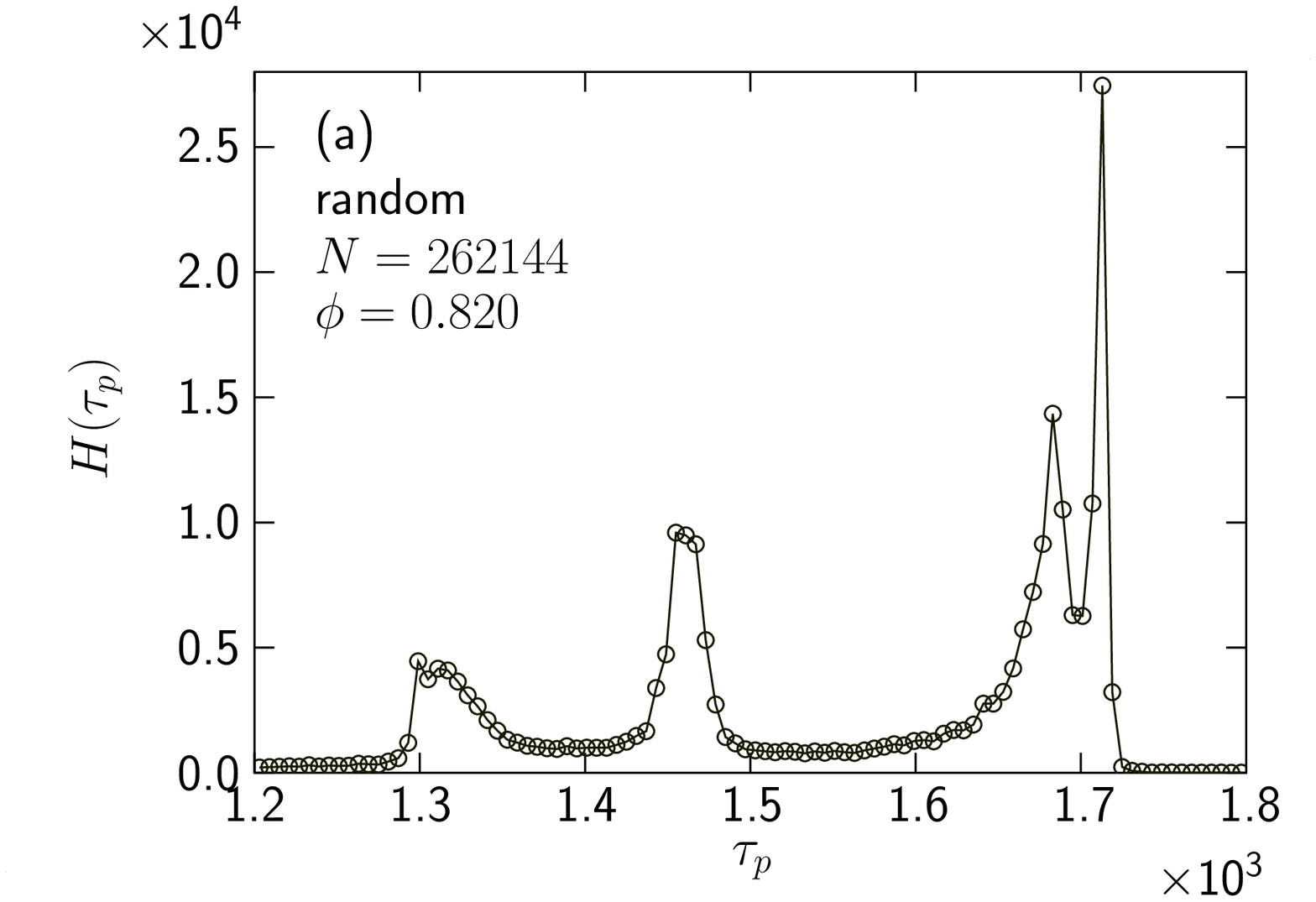}
  \includegraphics[width=6cm]{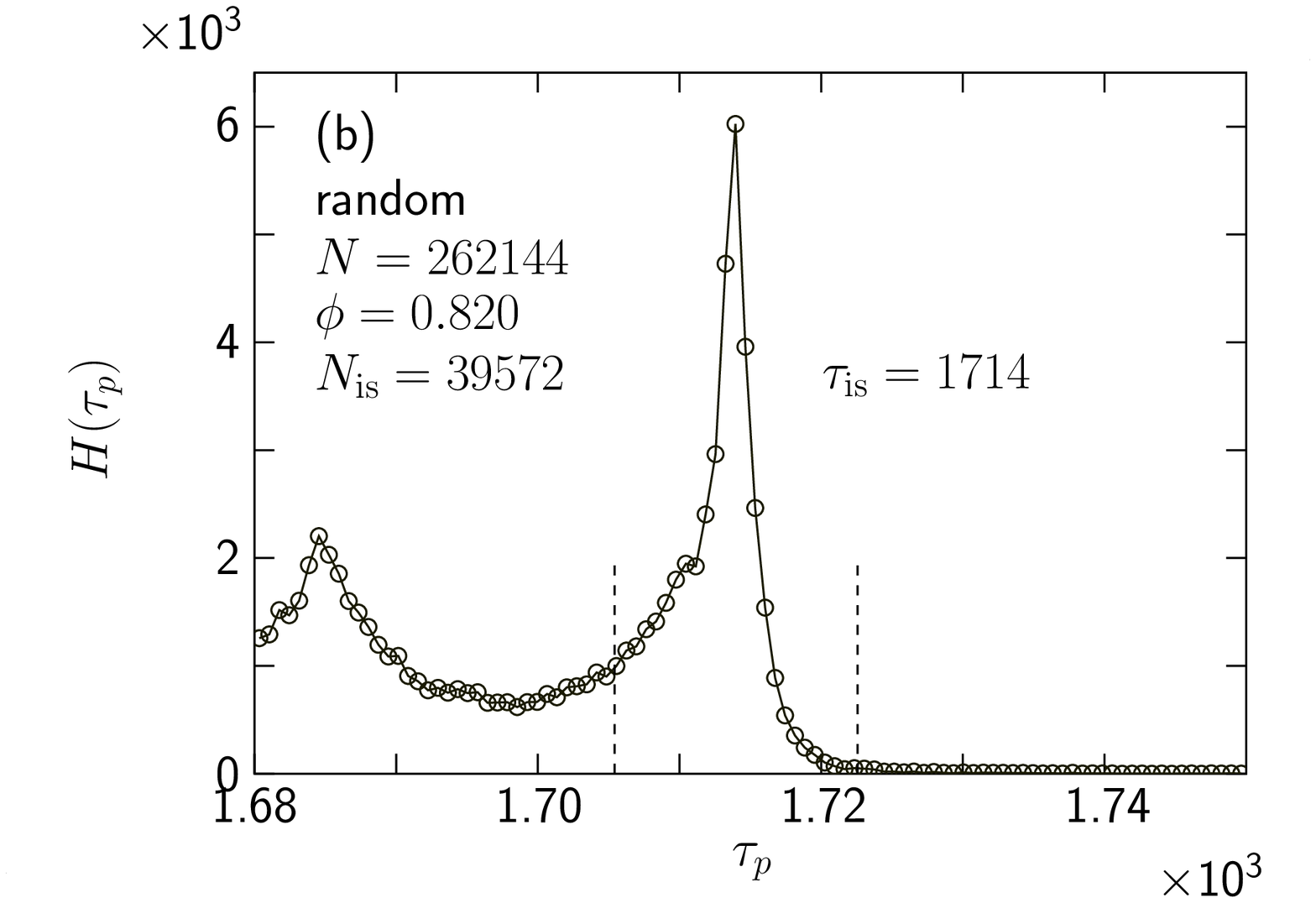}
  \caption{Determination of $\tau_\mathrm{is}$ and the size of the island,
    $N_\mathrm{is}$, which is the number of particles that belong to the corresponding
    peak in the histogram. Panel (a) shows the complete histogram and panel (b) is a zoom
    in on the uppermost peak. $N_\mathrm{is}$ is the number of particles with $\tau_p$ in
    the interval $\tau_\mathrm{is}\pm0.5\%$.}
  \label{fig:Htaup-island}
\end{figure}

For a better understanding of the NIB region---the large-$N$ region where the system
splits into different islands---we now examine the relation between relaxation time and
island size. It turns out that there is a direct relation between these quantities
such that large correlation times are possible only for big islands. We stress that the
results cannot be used to understand the properties outside the NIB region since these
systems experience collective relaxations that include all particles in the system, which
is a very different process from relaxations of isolated islands.

The determination of the key quantities---the number of particles in an island
$N_\mathrm{is}$ and its relaxation time $\tau_\mathrm{is}$---is illustrated in
\Fig{Htaup-island} where panel (a) shows the complete histogram $H(\tau_p)$ and panel (b)
zooms in on the uppermost peak. The relaxation time of the corresponding island is read
off from the figure and $N_\mathrm{is}$, the number of particles in the peak (and thus
also in the corresponding island), is determined from the interval with $\pm0.5\%$ around
the peak, as illustrated in the figure. For each configuration we here restrict our
analyses to the peak with the biggest relaxation time, to get the cleanest possible
analysis. For peaks at lower $\tau_p$ there is a higher risk that a peak could sometimes
be made up of data from two different islands, which would confuse the analysis.

\begin{figure}
  \includegraphics[width=7cm]{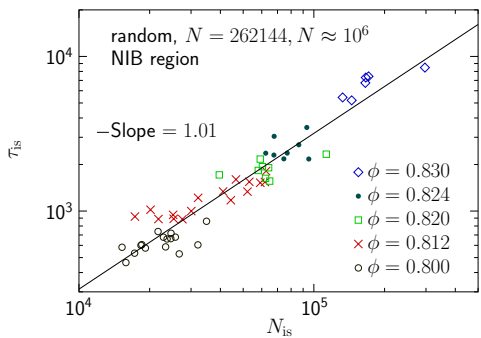}
  \caption{Relaxation time vs island size in the NIB region determined for the uppermost
    peaks as shown in \Fig{Htaup-island}(b). The slope $1.01$ strongly suggests a linear
    relation which is $\tau_\mathrm{is}\approx 0.029N_\mathrm{is}$.}
  \label{fig:tauis_Nis}
\end{figure}

The relation between island size and relaxation time is shown in \Fig{tauis_Nis}(a) for a
number of different configurations with $\phi=0.800$ through $0.830$, generated from
random configurations. Since the data give evidence for a linear behavior on a double-log
plot with slope $\approx1$, this points to a linear relation, and we find
$\tau_\mathrm{is}\approx 0.029 N_\mathrm{is}$.

We now suggest that this linear relation is an effect of the relaxation of a collection of
a set of partially overlapping particles. The relation between relaxation time and number
of particles in a one-dimensional model with $L$ particles of unit size, initially
compressed to have small overlaps, is found in \App{One-dimensional} to be $\tau\sim
L^2$. For the two-dimensional case, with the assumption that it is the lateral size that
determines the relaxation time, this translates into $\tau\approx 0.05 N$, which is the
same conclusion as from \Fig{tauis_Nis}, only with a somewhat bigger prefactor. A
difference in prefactor is not surprising since the disordered two-dimensional relaxed
systems, beside the different dimensionality, are very different from the simple,
perfectly ordered one-dimensional model considered in \App{One-dimensional}.

To make contact with the determination of $\tau$ in \Fig{tau-taup-N}(a) the relation
$\tau_\mathrm{is}\approx 0.029 N_\mathrm{is}$ is there shown as a dotted line. For each
$\tau$, that line shows the minimum $N_\mathrm{is}$ of an island with the relaxation time
equal to $\tau$. The fact that this line falls slightly to the left of the NIB line, is
consistent with the fact that a system may only have an island of size $N_\mathrm{is}$
if it consists of $N>N_\mathrm{is}$ particles.

It is also possible to construct a simple argument for the relation between density and
the average $N_\mathrm{is}$ which then also translates into the $\phi$-dependence of
$\taupg$.  For sufficiently big $N$ we expect a system at a density well below
$\phi_J^\mathrm{rnd}$ to be too dilute to make up a single well-connected island, and it
therefore breaks up into a set of weakly connected (or disjoint) islands with on the
average $N_\mathrm{is}(\phi)$ particles. If one simplifies greatly and considers the
islands to be made up of both a dense core with radius $R$ with
$\phi\approx\phi_J^\mathrm{rnd}$ and a thin buffer zone of thickness $\delta$, with no
particles, the total area associated with an island with radius $R$ is given by
$\pi R^2 + 2\pi R\delta$. By relating this area to the inverse density, and assuming that
there is no buffer zone at $\phi_J^\mathrm{rnd}$, the relevant expression becomes
\begin{displaymath}
  \frac{R^2 + 2R\delta}{R^2} = \frac{1/\phi}{1/\phi_J^\mathrm{rnd}}
  = \frac{\phi_J^\mathrm{rnd}}{\phi},
\end{displaymath}
and for densities close to $\phi_J^\mathrm{rnd}$ we get
\begin{displaymath}
  R \sim (\phi_J^\mathrm{rnd} - \phi)^{-1},\quad N_\mathrm{is}(\phi)
  \sim (\phi_J^\mathrm{rnd} - \phi)^{-2}.
\end{displaymath}
Together with $\tau_\mathrm{is}\sim N_\mathrm{is}$ this leads to
$\taupg\sim (\phi_J^\mathrm{rnd} - \phi)^{-2}$. For comparison we note that the divergence
of $\taupg$ in the NIB region from \Fig{tau-taup-N}(a) for $N\approx10^6$ and
$0.800\leq\phi\leq0.830$ gives the similar exponent $-2.1$. It could thus be that this
simple model captures the behavior well below jamming, but considering the bold
assumptions and simplifications involved here, there is no wonder that it fails to predict
the higher value of the exponent expected closer to jamming \cite{Olsson_Teitel:gdot-scale}.

\subsection{Analyses of $\Cforce$}

\begin{figure}
  \includegraphics[width=7cm]{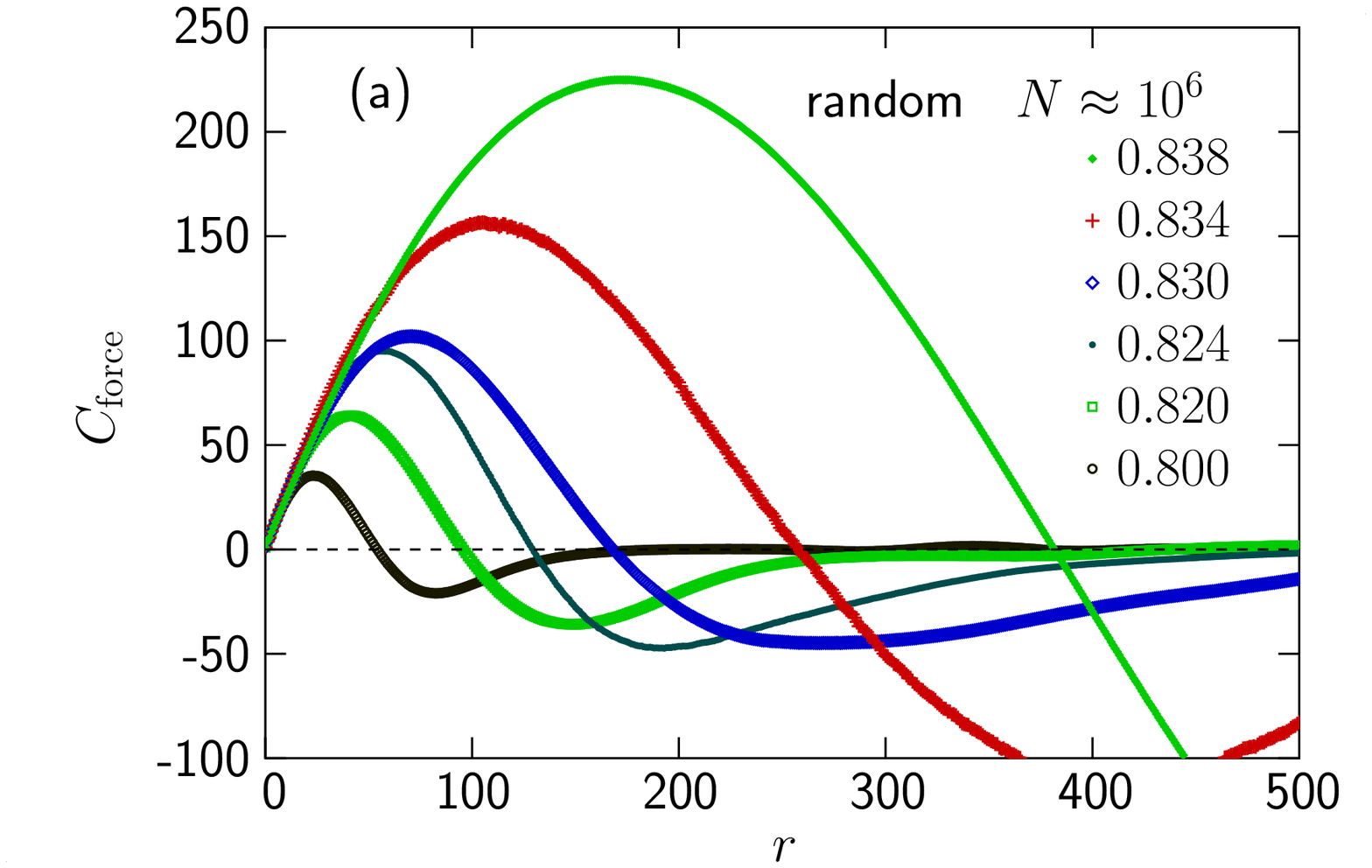}
  \includegraphics[width=7cm]{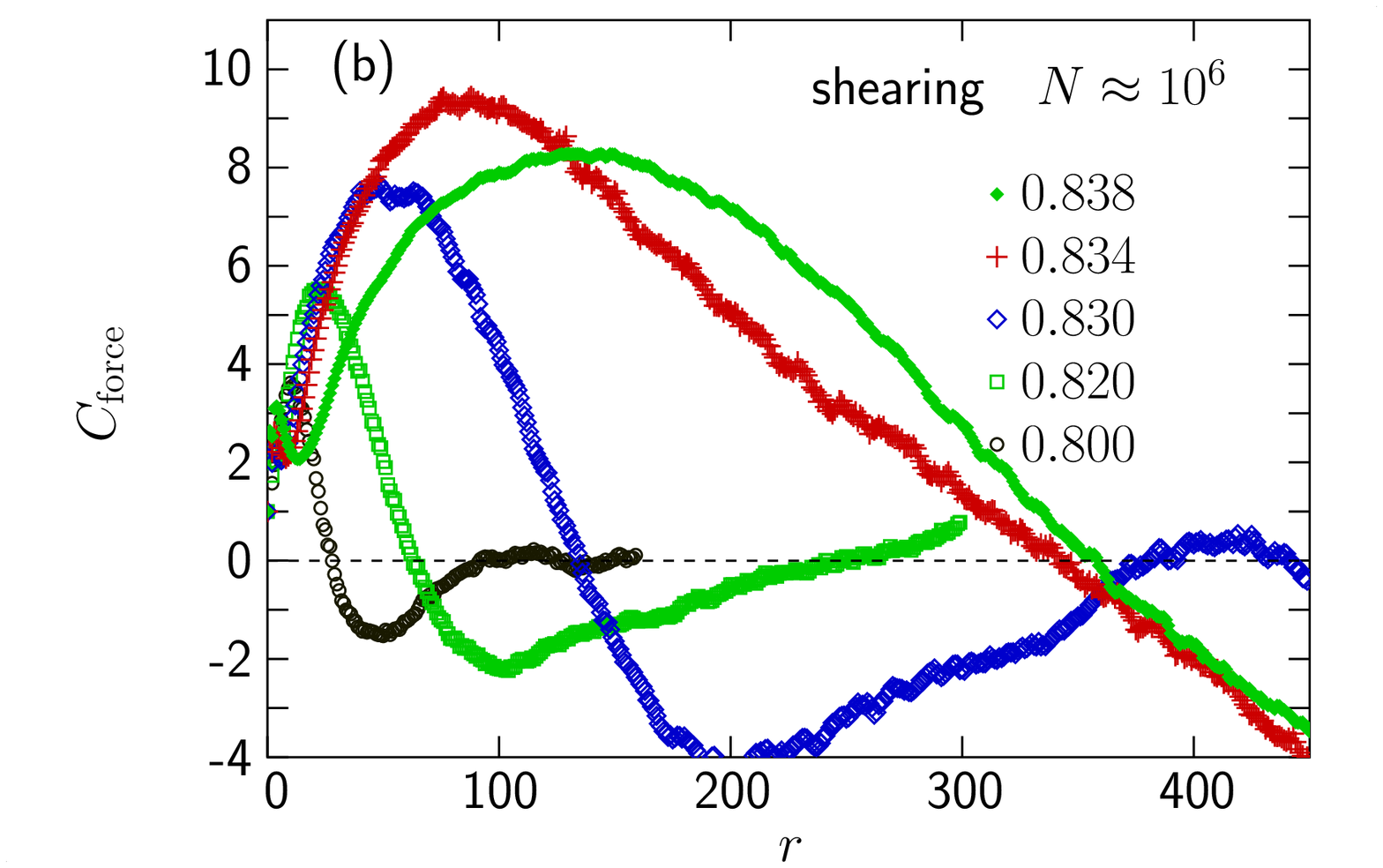}
  \caption{Force correlation function from \Eq{Cforce} for $N\approx10^6$ and $\phi=0.800$
    through 0.838. Panel (a) shows $\Cforce(r)$ for the random case whereas panel (b) is
    $\Cforce(r)$ for the shearing case. The magnitude of $\Cforce$ is always considerably
    bigger for the random case than for the shearing case, which is also why the data
    for the shearing case appear noisier. (In order not to clutter the noisier panel (b),
    we only show the data for the relevant region in $r$, somewhat past the crossing that
    determines $\xiforce$.) For the random case the height of the curves increases with
    increasing $\phi$ whereas for the shearing case the height appears to saturate or have
    a maximum at $\phi=0.834$.}
  \label{fig:Cforce}
\end{figure}

The correlation function $\Cforce(r)$ in \Eq{Cforce}, introduced in \REF{\NIB}, is shown
in \Fig{Cforce} for both the random and the shearing cases. We note that the magnitude of
$\Cforce(r)$ for the random case is much higher than for the shearing case and that the
weak signal in the latter case makes it difficult to get reliable data. Another difference
is that the height (the maximum value) of the curves for the random case keeps increasing
as $\phi$ increases whereas, in the shearing case, the height appears to saturate, or have
a maximum, at $\phi=0.834$.

\begin{figure}
  \includegraphics[width=7cm]{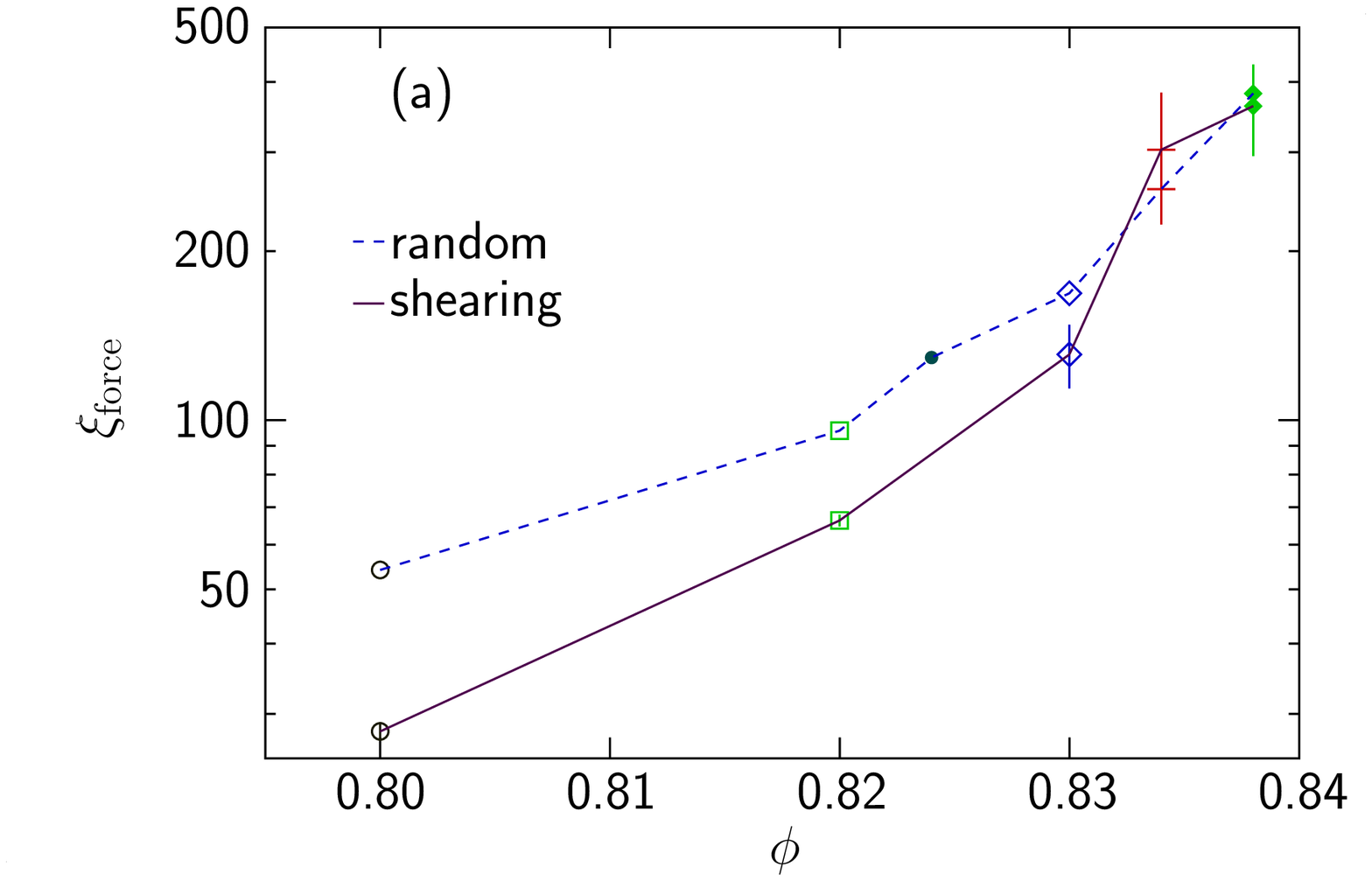}
  \includegraphics[width=7cm]{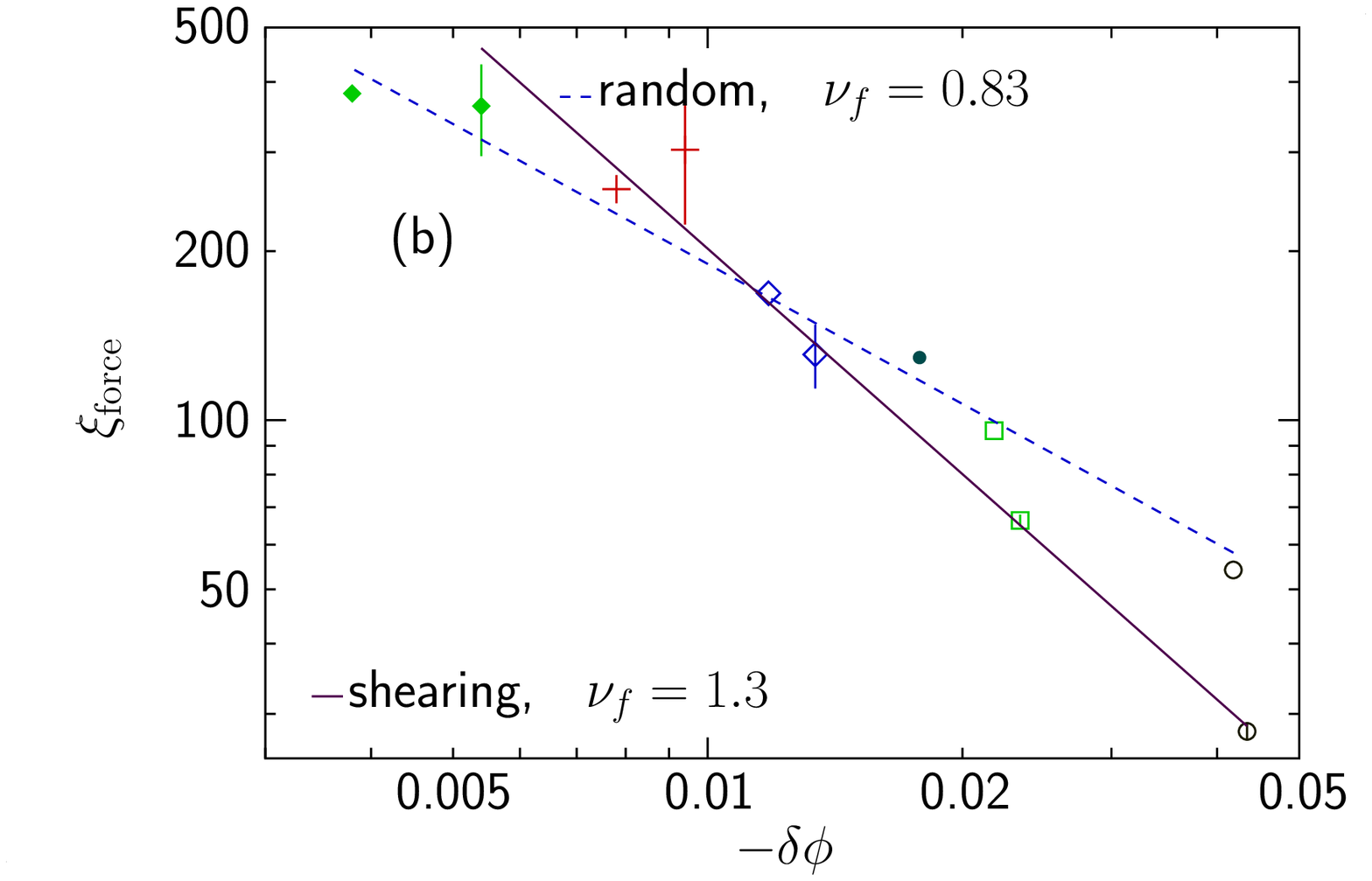}
  \caption{The length $\xi_\mathrm{force}$ for the two different protocols determined from
    relaxations with $N\approx10^6$. Panel (a) shows that $\xiforce$ behaves about the
    same for the two different protocols, but is always somewhat lower for the shearing
    case. Panel (b) shows attempted determinations of the critical exponent $\nu_f$ for
    the two cases, giving $\nu_f=0.92$ for the random case and $\nu_f=1.3$ for the
    shearing case.  We consider these values to be effective exponents since they describe
    the behavior well away from jamming.}
  \label{fig:xiforce}
\end{figure}

The length scale used to interpret the results in \REF{\NIB} is $\xiforce$, which is a
measure of the distance over which the net particle forces are correlated and is
determined from the distance where $\Cforce(r)$ turns negative.  \Figure{xiforce} shows
$\xiforce$ for the two different simulation protocols. The raw data in panel (a) show that
$\xiforce$ is consistently somewhat smaller in the shearing case than in the random case,
but also that this could possibly change at higher $\phi$. By fitting both sets of data to
algebraic divergences, $\xiforce\sim(-\delta\phi)^{-\nu_f}$ as shown in panel (b) we find
differing exponents, $\nu_f=0.92$ for the random case and $\nu_f=1.3$ for the shearing
case. We consider these values to be only effective exponents since they are determined
from $\xiforce$ well away from jamming.  (The determinations have been done with
$\delta\phi=\phi-\phi_J^\mathrm{rnd}$, with $\phi_J^\mathrm{rnd}=0.8418$
\cite{Vagberg_VMOT:jam-fss} for the random case, and $\delta\phi=\phi-\phi_J$, with
$\phi_J=0.8434$ \cite{Olsson_Teitel:gdot-scale}, for the shearing case.)

\begin{figure}
  \includegraphics[width=7cm]{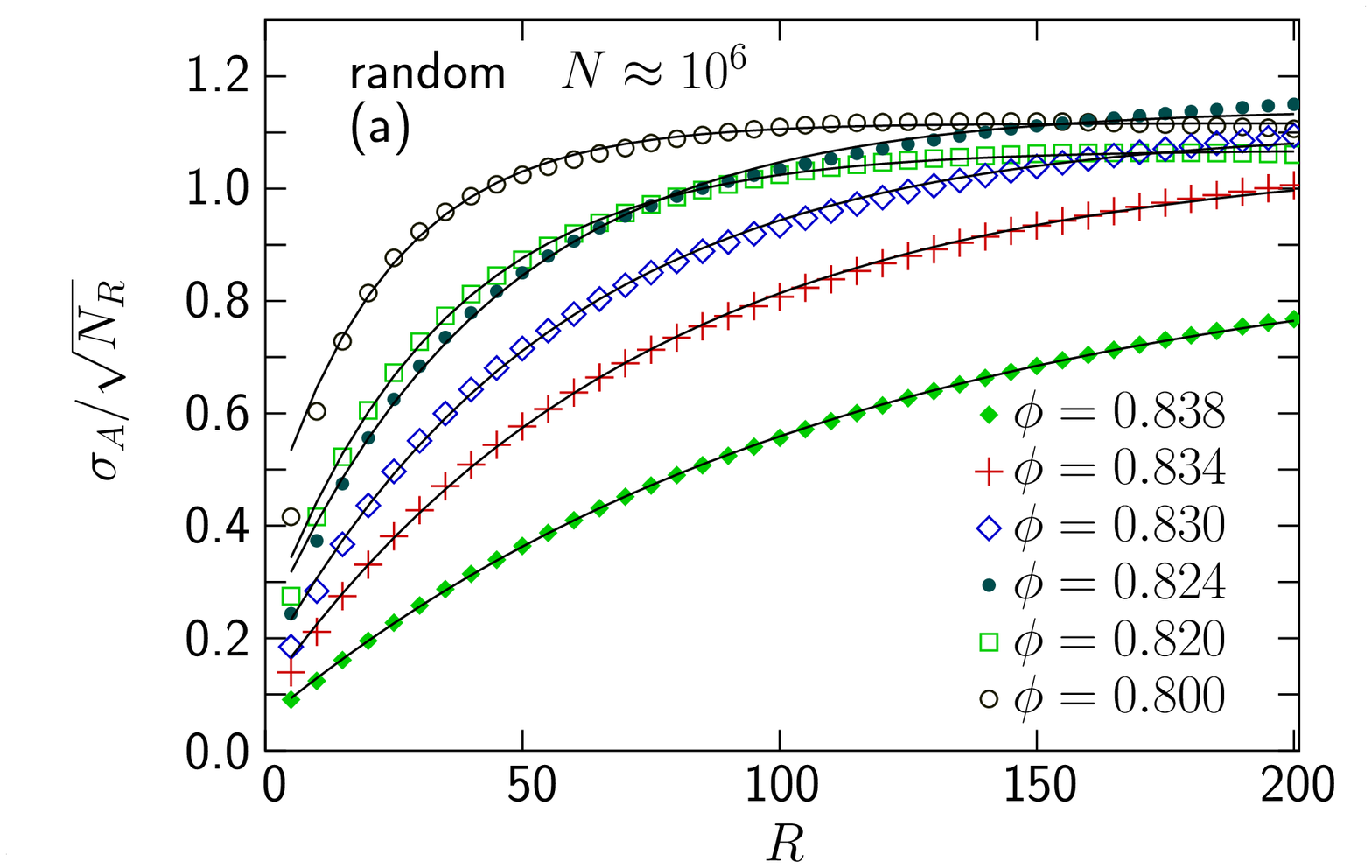}
  \includegraphics[width=7cm]{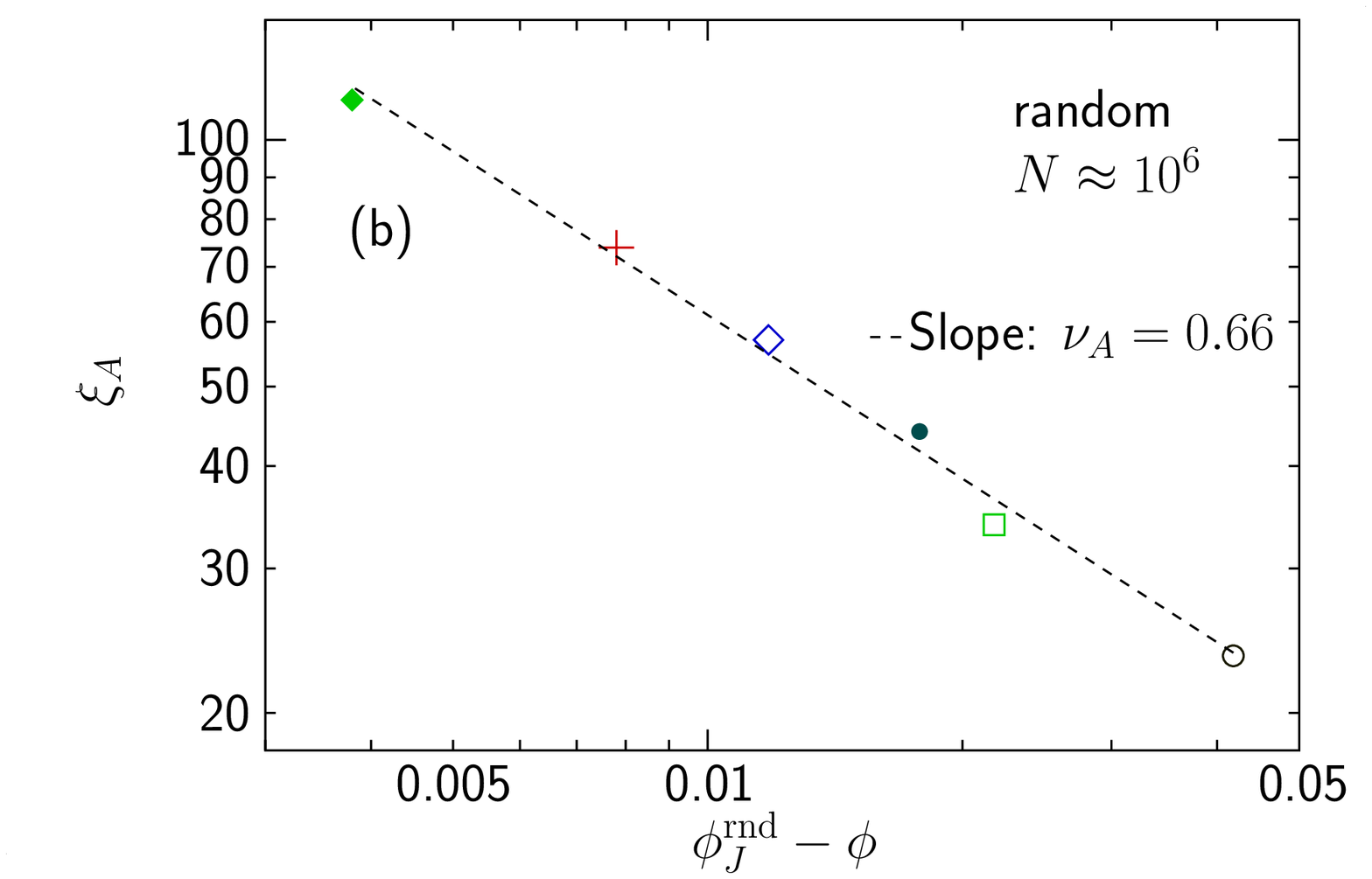}
  \caption{Determination of a characteristic length from the density fluctuations for the
    random case. Panel (a) shows density fluctuations for $N=262144$ and $\phi=0.800$
    through 0.838 together with lines from the fitting to \Eq{xiA}. Panel (b) is $\xi_A$
    from these fits vs $\phi^\mathrm{rnd}_J-\phi$. The fit of $\xi_A$ for $\phi\leq0.830$
    based on the assumption of an algebraic divergence, tentatively suggests $\nu_A=0.66$.}
  \label{fig:sigmaA}
\end{figure}

It is also possible to determine a length scale in the random case from the $R$-dependence
of the density fluctuations. To that end we determine $\sigma_A(R)$ from the relaxed
systems of $N\approx10^6$ particles for $\phi=0.800$ through 0.838 and fit data for
$R\geq15$ to
\begin{equation}
  \sigma_A(R)/\sqrt{N_R} = A_0(\phi) + A_1 e^{-R/\xi_A}.
  \label{eq:xiA}
\end{equation}
The data are shown in \Fig{sigmaA}(a). The solid lines are the fitted functions.  The
physical interpretation of the length scale $\xi_A$ is that the relaxation is only
effective in reducing the fluctuations at length scales smaller than $\xi_A$.
\Fig{sigmaA}(b) is $\xi_A$ vs $\phi_J^\mathrm{rnd}-\phi$. The assumption of an algebraic
divergence, $\xi_A\sim (\phi_J^\mathrm{rnd}-\phi)^{-\nu_A}$, with
$\phi_J^\mathrm{rnd}=0.8418$ \cite{Vagberg_VMOT:jam-fss}, gives $\nu_A=0.66$.

One could wonder whether our two different lengths, $\xiforce$ and $\xi_A$, really measure
the same thing but in different ways. The large difference of the exponents, $\nu_f=0.92$
and $\nu_A=0.66$, however, makes us conclude that that is most likely not the case,
especially since the two quantities are determined from the same sets of relaxed
configurations.

\section{Finite size effects in $\eta_p$}
\label{sec:etap}

\begin{figure}
  \includegraphics[width=8cm]{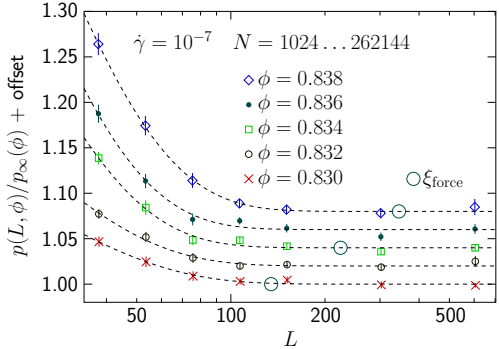}
  \caption{Finite size dependence of $p$ at $\phi=0.830$ through 0.838. In contrast to the
    prediction in \REF{\NIB} that the finite size effect is governed by the rapidly
    increasing $\xi_\mathrm{force}$ (marked by big open circles on the dashed curves for
    $\phi=0.830$, 0.834, and 0.838) the figure suggests an onset of finite size effects at
    $L\approx100$, independent of $\phi$.}
  \label{fig:etap}
\end{figure}

We now leave the discussion of the relaxation time determined from relaxations of two
different sets of initial configurations and instead turn to $\eta_p\equiv p/\gdot$
obtained from shearing simulations. What is in focus is the claim in \REF{\NIB} that
$\eta_p$ may not be used for extracting the critical behavior, since it is plagued by
strong finite size effects and is only reliable in the---presumably inaccessible---limit
$L\gg\xi_\mathrm{force}$. We do however note that there is no strong numerical data in
\REF{\NIB} in support of their claim and that our own data in \Fig{etap} gives no evidence
for the existence of any problematic finite size effect.

Before turning to our numerical data we shortly sketch what should be expected if the
jamming transition were working as an ordinary continuous transitions. (We find below that
it does not.) In shear driven systems below jamming there is ample evidence for a
correlation length $\xi\sim(\phi_J-\phi)^{-\nu}$, with $\nu\approx1$
\cite{Olsson_Teitel:jam-xi-ell}, and it is furthermore shown that the pressure equivalent
of the shear viscosity diverges as $\eta_p\sim\xi^{-\beta/\nu}$ with
$\beta/\nu\approx 2.7$. For quantities as pressure and shear stress one would then expect
the introduction of a finite (small) system size $L$ to mean that things should depend on
the dimensionless quantity $\xi/L$ and the expectation is then that the pressure should
start increasing as $L$ decreases below some $L_c(\phi)\sim\xi$. For the density range
$\phi=0.830$ through 0.838 considered below this $L_c$ should be expected to increase by a
factor of $\approx 2.5$ when the density increases from $\phi=0.830$ to 0.838.

To examine the finite size dependence \Fig{etap} we have determined $p(\phi,L)$ with shear
rate $\gdot=10^{-7}$ for $\phi=0.830$, 0.832, 0.834, 0.836, and 0.838 and $N=1024$ through
262144 particles, corresponding to system sizes $L\approx37$ through $\approx 600$ . The
data are shown in \Fig{etap} as $p(\phi,L)/p_\infty(\phi)$ vs $L$. Since the data are
normalized to unity for large $L$, the points for different $\phi$ are shifted vertically
for clarity. The extrapolated $p_\infty(\phi)$ are obtained by fitting
\begin{equation}
  \label{eq:pL}
  p(\phi,L) = p_\infty(\phi) + C_p(\phi) e^{-L/L_c},
\end{equation}
shown by the dashed curves in \Fig{etap}. 

In the figure we find evidence for a clear finite size effects setting in below a constant
$L\approx100$, for all the different densities. Likewise, from the fit to \Eq{pL} we find
$L_c\approx 21$ for all the densities without any trend. This is thus at odds with the
expectation that $p$ should depend on $L/\xi(\phi)$ and is difficult to reconcile with the
common framework for shear driven jamming. The invocation of logarithmic corrections to
scaling---instead suggesting a scaling with $L(\ln L)^{1/4}/\xi(\phi)$
\cite{Kenna:ucd}---does not resolve the issue. For now we just comment that this
points to the need for a novel picture of shear-driven jamming. Such work is in progress
and will be presented elsewhere.

\section{Summary}
\label{sec:Discuss}

We have performed a thorough study of the relaxation dynamics in a 2D model of jamming
with two different simulation protocols: relaxation of random configuration and
relaxation of configurations generated through steady shearing. The question in focus has
been on finite size effects and the possibility to reliably determine the critical
behavior with two different methods (1) from the relation between the relaxation time,
$\tau$ and the contact number deficiency, $\delta z$, and (2) from the pressure equivalent
of the shear viscosity.

For the first point we first note that the identification of the NIB effect in \REF{\NIB},
together with the claim that it works about the same for both the random case and the
shearing case appears correct. Our crucial observation is, however, that there is also
another finite size effect---the DF (density fluctuations) effect---which is only clearly
seen for the random case and has profound consequences and adversely affects the data
produced with that protocol. This then gives a rational for trusting the analyses obtained
with the shearing protocol in spite of the presence of problematic finite size effects in
data obtained with the other protocol as in Fig.~1 of \REF{\NIB}.

Another key result is that it is possible to determine the average particle relaxation
time $\taupg$, which is a quantity with a well defined thermodynamic limit, which
therefore may be used to define the critical divergence, but also that $\tau$---the energy
relaxation time which has a problematic $\ln N$-dependence for very large $N$---may
be used as a reliable estimate of $\taupg$ if one just stays clear from the very large $N$
region of the NIB effect.

When it comes to the second point on the finite size effect on $\eta_p$, our data for
$0.800\leq\phi\leq0.838$ gives evidence for significant finite size effects for sizes
below $L\approx100$, independent of $\phi$. We note first that this suggests that the data
for $N=65536$ and $L\approx300$ used in scaling analyses of shear driven jamming
\cite{Olsson_Teitel:gdot-scale} are not too much affected by finite size effects, but also
that this $\phi$-independent finite size effect is at odds with the expectations from
shear-driven jamming being an ordinary continuous transition.

To summarize we have examined the evidence behind the criticism of two different methods
employed to determine the critical divergence of shear-driven jamming, and find that the
conclusions were unnecessarily pessimistic. There do nevertheless remain many questions
regarding reliable and consistent interpretations of such data as well as regarding the
validity of the related theoretical approaches \cite{DeGiuli:2015,
  Harukuni-logcorr:2020, Olsson:jam-vhist}.

I thank Y. Nishikawa, A. Ikeda, L. Berthier for useful discussions and S. Teitel for many
discussions as well as comments on a previous version of the article.  The computations
were enabled by resources provided by the Swedish National Infrastructure for Computing
(SNIC), partially funded by the Swedish Research Council through grant agreement no.\
2018-05973.

\appendix
\section{Evidence for this extra finite size effect from data in \REF{\NIB}}
\label{sec:NIB-evidence}

We here point out that the conclusion of the existence of a finite size effect other than
the NIB effect, here arrived at through analyses of $H(\tau_p)$, may actually also be reached
from the information presented in \REF{\NIB}.

In \Sec{Comp_ran_shear} we argued that $N=16384$ at $\phi=0.820$ is actually outside the
NIB region and that the finite-size dependence there therefore has to have another origin
than the NIB effect. From the statement \cite{\NIB} that the small-$N$ regime is a regime
where ``the force correlation is limited by the system size, so that
$\xiforce(\phi,N)/L=O(1)$'', together with their Fig.~5(b) which shows that
$\xiforce\sim N^{1/2}$ at least up to $N=128^2$, it seems that $N=16384\equiv128^2$ at
$\phi=0.820$ is actually in this small-$N$ regime. Furthermore, their Fig.~4(c) which shows
the coarse-grained force field in a big system at the same density, $\phi=0.820$, has
structures bigger than $L\approx150$, which is the linear size of a system with $N=16384$.

Similarly, with regards to their Fig.~4(d) they write ``a single correlated island seems
to cover the whole system for $\phi=0.83$, despite the fact that the system size is large
$N = 262144$''. This therefore suggests---in agreement with our \Fig{tau-taup-N}(a)---that
this point is outside the NIB region which also implies that the $N$-dependence at
$\phi=0.830$ in their Fig.~2(a), which shows data for $N\leq262144$, needs an explanation
other than the NIB effect.

\section{Size-dependence from the binomial distribution}
\label{sec:binomial}

For a detailed comparison of $\sigma_A$ with $\sigma_{N_R}$ one has to take two different
factors into account. The first is the presence of two different particle sizes and the
second is the fact that for particles on the circle, only the fraction of the particles
within the circle contributes to $A(R)$.

We first introduce $A_c$ for which the particles with their centers inside $R$ are instead
included with their total area. With the areas of the big and small particles denoted by
$a_b$ and $a_s$, the variance in $A_c$ gets contributions from two different sources with,
on the average $N_R/2=fN/2$ particles each, giving
\begin{equation}
  \sigma_{A_c}^2 = \sigma_{A_c,b}^2 + \sigma_{A_c,s}^2 = (a_b^2+a_s^2) f(1-f)\frac N 2 =
  \expt{a^2} f(1-f)N,
\end{equation}
which gives
\begin{equation}
  \label{eq:sigmaAN}
  \sigma^2_{A_c}/N_R = \expt{a^2}(1-f).
\end{equation}

Most of the deviations of $\sigma_A/\sqrt{N_R}$ from the analytically expected results in
\Fig{saa-8200}(a) are due to the difficulty in getting good precision in fluctuation
quantities. There are however clear deviations from the analytical curves at small $R$
which cannot be attributed to the limited statistics.  It turns out that this is related
to the way the particles that are only partly inside the region are handled. It is then only
the part of the area that is inside the radius $R$ that is counted. For small $R$ a bigger
fraction of the particles are at the boundaries and it appears that this has the effect to
reduce the fluctuations.

To compare with expressions for the variance of the density we note that the density is
given by $\phi_R = A(R)/(\pi R^2)$ and that the variance of the density therefore becomes 
\begin{displaymath}
  \mathrm{var}(\phi_R) = \frac{\sigma_A^2}{(\pi R^2)^2}
  = \frac{1}{\expt a \pi R^2}\frac{\sigma_A^2}{N_R}.
\end{displaymath}

\section{One dimensional model}
\label{sec:One-dimensional}

To understand the relation between $N_\mathrm{is}$ and $\tau_\mathrm{is}$ we turn to a
one dimensional model with $L$ particles with diameter equal to unity. The starting point
is the zero-force configuration
\begin{displaymath}
  x_i = i-(L-1)/2,\quad i=0,\ldots L-1,
\end{displaymath}
which is compressed to
\begin{displaymath}
  r_i = x_i + \eta_i,
\end{displaymath}
such that the overlap between neighboring particles becomes
\begin{displaymath}
  \delta_{i,i+1} = 1 - (r_{i+1}-r_i) = \eta_i - \eta_{i+1}.
\end{displaymath}
With force $f_{i,i+1} =\epsilon \delta_{i,i+1}$ the total force on particle $i$, not at
one of the boundaries, is
\begin{displaymath}
  f_i = \epsilon \delta_{i-1,i} - \epsilon \delta_{i+1,i} =
  \epsilon(\eta_{i+1} - 2\eta_{i} + \eta_{i-1}) \approx \epsilon\Delta^2\eta,
\end{displaymath}
and by assuming overdamped dynamics, $dr_i/dt = f_i/k_d$ one arrives at
\begin{displaymath}
  \frac{\partial\eta}{\partial t} = \frac{\epsilon}{k_d} \Delta^2\eta.
\end{displaymath}
Going to the continuum this becomes a partial differential equation
\begin{displaymath}
  \frac{\partial\eta}{\partial t} = \frac{\epsilon}{k_d} \frac{\partial^2\eta}{\partial x^2}.
\end{displaymath}
With zero-force boundary conditions at $x=0$ and $x=L$ the solution becomes
\begin{displaymath}
  \eta(t,x) = -A e^{-t/\tau_\eta} \sin(\pi x/L),
\end{displaymath}
which gives the relaxation time
\begin{displaymath}
  \tau_\eta = \frac{k_d}{\epsilon}\left(\frac L\pi\right)^2,
\end{displaymath}
and by finally considering the decay of the energy, with is $\sim\delta^2$, one arrives at 
\begin{displaymath}
  \tau_\mathrm{1d}/(k_d/\epsilon) = \frac 1{2\pi^2} L^2 \approx 0.05 L^2.
\end{displaymath}

When the particles are instead arranged in an orderly two dimensional array (which is of
course very different from the situation in our simulations) we make use of $L=\sqrt N$ to
get $\tau_\mathrm{2d}(N)/(k_d/\epsilon)\approx0.05 N$.

%

\begin{thebibliography}{31}
\bibitem [{Cfo()}]{Cforce}%
  \BibitemOpen
  \href@noop {} {}\bibinfo {note} {The quantity plotted in \REF{\NIB} is
  obtained by also dividing by $2\pi r$ to compensate for the contribution of
  the increasing number of particles with the distance. We have made use of
  that expression without this division to make it easier to compare the shape
  of these curves.}\BibitemShut {Stop}%
\end{thebibliography}

\end{document}